\documentclass[aps,pre,twocolumn,superscriptaddress,floatfix,fleqn,10pt,showpacs]{revtex4-1}
\usepackage{graphicx}
\usepackage{verbatim}
\usepackage{comment}
\usepackage{amssymb}
\usepackage{stmaryrd}
\usepackage{enumerate}
\usepackage{color}
\usepackage{hyperref}

\newcommand{\disp}[1]{Eq.~(\ref{#1})}
\newcommand{\refdisp}[1]{Ref.~(\onlinecite{#1})}
\newcommand{\figdisp}[1]{Fig.~(\ref{#1})}

\newcommand{\lsim}{\mathrel{\hbox{\rlap{\lower.55ex \hbox{$\sim$}} \kern-.3em \raise.4ex \hbox{$<$}}}}

\begin{document}

\def\tj{{$t$-$J$~}}
\def\Pd{{Pad{\'e} }}

\title{Linked-cluster expansion for the Green's function of the infinite-$U$ Hubbard model}

\author{Ehsan Khatami}
\affiliation{Department of Physics, University of California, Santa Cruz, CA 95064, USA}
\affiliation{Department of Physics, University of California, Davis, CA 95616, USA}
\author{Edward Perepelitsky}
\affiliation{Department of Physics, University of California, Santa Cruz, CA 95064, USA}
\author{Marcos Rigol}
\affiliation{Department of Physics, The Pennsylvania State University, University Park, PA 16802, USA}
\author{B. Sriram Shastry}
\affiliation{Department of Physics, University of California, Santa Cruz, CA 95064, USA}


\begin{abstract}
We implement a highly efficient strong-coupling expansion for the Green's
function of the Hubbard model. In the limit of extreme correlations, where the
onsite interaction is infinite, the evaluation of diagrams simplifies
dramatically enabling us to carry out the expansion to the eighth order in
powers of the hopping amplitude. We compute the finite-temperature Green's 
function analytically in the momentum and Matsubara frequency 
space as a function of the electron density. Employing Pad\'{e} approximations,
we study the equation of state, Kelvin thermopower, momentum distribution 
function, quasiparticle fraction, and quasiparticle lifetime of the system at 
temperatures lower than, or of the order of, the hopping amplitude.  
We also discuss several different approaches for obtaining the spectral 
functions through analytic continuation of the imaginary frequency Green's function,
and show results for the system near half filling.
We benchmark our results for the equation of state against those obtained 
from a numerical linked-cluster expansion carried out to the eleventh order.
\end{abstract}

\maketitle

\section{Introduction}

In 1991, Metzner put forth an algorithm to compute the finite-temperature Green's 
function of the Fermi-Hubbard model [\disp{eq:hamiltonian}] through a linked-cluster 
strong-coupling expansion~\cite{w_metzner_91}. His approach offers a relatively
straightforward implementation on a computer, which is particularly useful today given 
the enormous improvements in computer power in the past two decades.
The Metzner formalism further simplifies in the limit of extreme correlations, as the 
onsite repulsion, $U$, tends to infinity. In this paper, we implement his approach to 
obtain analytical expressions for the single-particle Green's function in that limit through eighth 
order in the expansion parameter $\beta t$, where $\beta$ is the inverse temperature 
and $t$ is the hopping amplitude of the electrons on the lattice. 

In another  recent development, the  extremely correlated Fermi liquid theory (ECFL) \cite{ecfl} 
addresses this important limit through the use of Schwinger's source formulation of 
field theory. One of the significant physical ideas to come out of this theory is the 
presence of particle-hole asymmetry in the spectral densities of the single-particle 
Green's function and the Dyson-Mori self-energy \cite{ecfl, s_shastry_11, Asymm, 
hansen-shastry,ECFLAM,ECFLDMFT, larged}. This asymmetry, which has also been 
observed in dynamical mean-field theory studies of the Hubbard model~\cite{ECFLDMFT,Dengetal,DMFTrev},
becomes more pronounced as the density approaches half-filling, i.~e., as $n \to 1$. 
The asymmetry  has implications for understanding the magnitude and sign of the  
Seebeck coefficient near the Mott insulating limit \cite{Palsson,louis,Dengetal} and for 
 explaining the anomalous line shapes of angle-resolved photoemission spectroscopy 
 experiments \cite{gweon} in strongly correlated materials.

In a recent work \cite{e_khatami_13}, the present authors (with Hansen)  
used the series expansion method  to successfully  benchmark the  ECFL results 
for the spectral function \cite{hansen-shastry}, in their  common regime of  applicability.  
The currently available [$O(\lambda^2)$] self-consistent  solution of the ECFL   
is valid for  $n\lsim 0.75$. Additionally,  the 
insight afforded by the  aforementioned particle-hole asymmetry enabled us to construct a suitably 
modified first moment of the spectral function,  providing a good estimate for the location of the 
quasiparticle peak.  This moment reduces  the contribution from the  occupied side of the spectrum 
relative to the  unoccupied side, leading to a sharper location of the peaks. Therefore, using the series 
expansion to calculate this moment, we were able to study the dispersion of the quasiparticle energy 
and, as a result, the evolution of the Fermi surface in the limit $n\to 1$, i.e., beyond the density regime
currently accessible to the $O(\lambda^2)$ version of the ECFL.

Here, we expand upon our previous findings and perform analytic continuation to obtain
the full spectral functions. Direct analytic continuation of finite series, however, leads to 
unphysical results,  e.g., negative spectral functions can arise due to the truncation of 
the series. This is a well-studied problem with known resolutions~\cite{moments,Pairault}.
Therefore, and in particular, to ensure the positivity of spectral 
densities, we either take advantage of a transformation that guarantees this positivity, 
or assume an approximate form for the spectral functions, which comes out of the ECFL. We find a
good agreement between results from the two approaches, which capture the expected 
features of the spectra discussed above.

Using strong-coupling expansions, there have been several earlier studies of the 
thermodynamics and time-independent correlations of the Hubbard and related models 
\cite{seriesexpansions}. However, strong-coupling expansions for the 
{\em time-dependent correlations} are rare \cite{Pairault,self-consistent,l_craco}. In 
Ref.~\cite{Pairault}, the authors carried out a strong-coupling expansion 
for the Green's function to fifth order in $\beta t$ for the finite-$U$ Hubbard model. Here, the 
simplifications arising from the $U\to\infty$ limit allow us to go to eighth order in $\beta t$.
This provides us with the opportunity to employ \Pd approximations and study several 
static and dynamic quantities, such as the equation of state, momentum distribution function, 
the quasiparticle fraction, and lifetime at temperatures lower than the hopping amplitude, 
where the direct sums in the series do not converge. We also take advantage of the 
state-of-the-art numerical linked-cluster expansions (NLCEs)~\cite{NLCE-tJ}, developed 
recently for the \tj model, and set the exchange interaction $J$ to 0, to gauge our 
low-temperature equation of state obtained from the \Pd approximations. 

The organization of the paper is as follows: In Secs. \ref{model} and \ref{implementation},
we review the Metzner formalism and 
detail its numerical implementation. In Sec. \ref{GreenSigma}, we provide analytical 
expressions for the Green's function and the Dyson-Mori self-energy in momentum and 
Matsubara frequency space as a function of the density. In Sec. \ref{Pade}, we discuss 
the convergence of the series both before and after the use of Pad\'{e} approximations. 
Additionally, using the series, we report results for the time-dependent local Green's 
function, the equation of state, Kelvin thermopower, the quasiparticle weight at the 
Fermi surface, momentum occupation number, and quasiparticle lifetime and spectral functions at 
different points along the irreducible wedge of the Brillouin zone. We summarize our results
in Sec. \ref{sec:summ}.

\section{The model \label{model}}

In the strong-coupling limit, the Hubbard Hamiltonian is written as the sum of the unperturbed 
local Hamiltonian $H_0$, and a perturbation $H_1$ that accounts for hopping of electrons 
between the  sites of the lattice,
\begin{equation}
H=H_0+H_1,
\label{eq:hamiltonian}
\end{equation}
where
\begin{eqnarray}
H_0&=&U\sum_i{n_{i\uparrow}n_{i\downarrow}}-\mu\sum_{i\sigma}n_{i\sigma}
\nonumber \\
H_1&=&-\sum_{ij\sigma}t_{ij}c^{\dagger}_{i\sigma}c_{j\sigma}.
\label{Hubbard Hamiltonian}
\end{eqnarray}
Here, $c_{i\sigma}$ ($c^{\dagger}_{i\sigma}$) annihilates (creates) a
fermion with spin $\sigma$ on site $i$, $n_{i\sigma}=c^{\dagger}_{i\sigma}
c_{i\sigma}$ is the number operator, $U$ is the onsite repulsive Coulomb 
interaction, $\mu$ is the chemical potential,  
and $t_{ij}$ is the hopping matrix element between sites $i$ and $j$. As
discussed in the following, we allow for nearest-neighbor 
hopping only, namely, $t_{ij}=t$ if $i$ and $j$ are nearest neighbors, 
and $t_{ij}=0$ otherwise. 

\section{Metzner's approach for computing the Green's function \label{implementation}}

We start by describing the Metzner formalism before turning our focus to topics related to 
its computational implementation in the limit of extreme 
correlations. Following the conventions in Ref.~\cite{w_metzner_91}, we define the 
finite-temperature single-particle Green's function as
\begin{equation}
\label{eq:g}
G_{\sigma j j'}(\tau-\tau')=-\left<T_{\tau}c_{j\sigma}(\tau)
c^{\dagger}_{j'\sigma}(\tau')\right>,
\end{equation}
where $\left<..\right>$ denotes the thermal average with respect to $H$,
$T_{\tau}$ denotes the imaginary time-ordering operator,
and the creation and annihilation operators in the Heisenberg representation are
expressed as
\begin{eqnarray}
c^{\dagger}_{j\sigma}(\tau)&=&e^{H\tau}c^{\dagger}_{j\sigma}e^{-H\tau}, \nonumber
\\
c_{j\sigma}(\tau)&=&e^{H\tau}c_{j\sigma}e^{-H\tau},
\end{eqnarray}
where $0\leq\tau\leq\beta$ is an imaginary time variable.

To derive a perturbative expansion for $G_{\sigma j j'}(\tau-\tau')$, we switch
to the interaction representation,  where the time evolution of the operators is governed by 
the unperturbed Hamiltonian, $H_0$. The Green's function can then be expressed as
\begin{equation}
\label{eq:G0}
G_{\sigma j j'}(\tau-\tau')=-\left<T_{\tau}c_{j\sigma}(\tau)
c^{\dagger}_{j'\sigma}(\tau')\mathcal{S}\right>_0/\left<\mathcal{S}\right>_0,
\end{equation}
where the expectation values ($\left<..\right>_0$) are taken with respect to the unperturbed 
Hamiltonian, and $\mathcal{S}$ is given by
\begin{equation}
\label{eq:S}
\mathcal{S}=T_{\tau}\textrm{exp}\left[\int_0^{\beta}d\tau \sum_{ij\sigma}t_{ij}
c^{\dagger}_{i\sigma}(\tau)c_{j\sigma}(\tau)\right].
\end{equation}
Next, by expanding the exponential in Eq.~(\ref{eq:S}), both the numerator and the 
denominator of Eq.~(\ref{eq:G0}) can be written as perturbative series expansions
in $t$. As detailed in Ref.~\cite{w_metzner_91}, every term of the expansions
can be written in terms of cumulants (connected many-particle Green's functions) of 
the unperturbed system, denoted by $C^0_m$ ($m$ indicates the number of creation 
or destruction operators in the cumulant). Due to the local nature of the unperturbed 
Hamiltonian, the cumulants are site diagonal, i.e., the only nonzero ones are those 
whose site variables are the same, and they can therefore be indexed by site. Due to 
the translational invariance of the Hamiltonian, an order $m$ cumulant at site $i$ is 
independent of $i$ and is a function of only the time and spin indices of the $m$ creation, 
and $m$ destruction operators acting on $i$, i.e., $C^0_{mi}\equiv C^0_{m}
(\tau_1\sigma_1, \dots \tau_m\sigma_m|\tau'_1\sigma'_1, \dots \tau'_m\sigma'_m)$.
As we will see in the following, this invariance is a major advantage of the present method. 
Using it,  each term in the expansion can be written as a product of a spatial 
part and a temporal part, which may then  be evaluated independently.

The terms in the expansion for $\left<\mathcal{S}\right>_0$ can be
evaluated using a diagrammatic approach, where each diagram consists of 
vertices, and directed lines connecting the vertices. Each vertex represents a site 
on the lattice, and each line represents a hopping process between two sites.
Furthermore, the spatial sums reduce to calculating free multiplicities of 
graphs when embedded on the lattice. This  computationally inexpensive part of 
the algorithm can be carried out independently of the most 
expensive part (taking the time integrals), for {\em any} lattice geometry.

The expectation value in the numerator of Eq.~(\ref{eq:G0}) can be
calculated the same way as $\left<\mathcal{S}\right>_0$, except that any 
graph in the former contains two additional {\em external} lines, one entering 
the site $j'$ at time $\tau'$ and one exiting the site $j$ at time $\tau$. Consequently,
in the lattice sums, one has to ``fix'' the vertices to which the external lines 
attach to be the sites $j$ and $j'$ on the lattice with the desired separation 
between them.

Another important feature of this method is the fact that the spatial sums 
are unrestricted (different vertices are allowed to be on the same lattice site), 
and therefore it can be verified that the contributions of disconnected diagrams 
are products of the contributions of their connected components (the linked-cluster 
theorem holds). Hence, the disconnected diagrams in the numerator of Eq.~(\ref{eq:G0}) 
are canceled by the denominator, and $G_{\sigma jj'}(\tau,\tau')$ is given as the sum 
of the contributions of only the connected graphs in its numerator.

Further details of the method are given in Ref.~\cite{w_metzner_91} and will not
be repeated here. The rules mentioned in Ref.~\cite{w_metzner_91} for generating the 
graphs and evaluating their contributions are reproduced below.

\subsection{Rules for calculating the one-particle Green's function
diagrammatically}
\label{sec:steps}

(i) Draw all topologically distinct diagrams: vertices connected by directed
lines such that the number of entering and exiting lines at each vertex is the same. 
The graphs consist of the internal lines that connect two vertices as well as two 
{\em external} lines that enter a vertex and exit a vertex. The order to which each 
graph contributes is equal to the number of internal lines it has. \\

(ii) Label each line with a time and spin index, and each vertex by a lattice
index. The vertex that has the entering external line is labeled by $j'$ and the 
vertex that has the exiting external line is labeled by $j$.\\

\begin{figure}[t]
\includegraphics[width=2.5in]{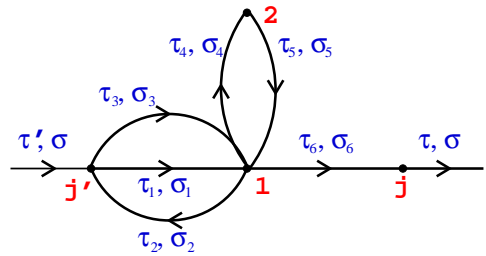}
\caption{Diagram of a sample graph in the sixth order (with six internal 
and two external lines). The time and spin indices of lines are ordered 
according to rule (iii). To calculate the contribution of this graph, we 
need to insert $C^0_2$, $C^0_3$, $C^0_1$, and 
$C^0_1$ for vertices $j'$, 1, 2, and $j$, respectively, for the time integral and
the spin sum, and $(t_{j'1})^3(t_{12})^2t_{1j}$ for the spatial sum. The symmetry
factor is 2 since exchanging lines that correspond to $\tau_1\sigma_1$ and 
$\tau_3\sigma_3$ does not change the topology of the graph.}
\label{fig:graph}
\end{figure}

(iii) Order the lines by defining a path that starts from the entering external
line at vertex $j'$, goes through all of the vertices, and ends with the exiting 
external line at $j$. Figure \ref{fig:graph} shows an example of such a graph in 
the sixth order.\\

(iv) Insert a factor of $(-t_{il})$ for each line that connects vertex $i$ to vertex $l$.\\

(v) Insert $C^0_{m}(\tau_1\sigma_1, \dots \tau_m\sigma_m|\tau'_1\sigma'_1, \dots
\tau'_m\sigma'_m)$
for each vertex that has $m$ entering lines labeled $\tau'_1\sigma'_1, \dots
\tau'_m\sigma'_m$ 
and $m$ exiting lines labeled $\tau_1\sigma_1, \dots \tau_m\sigma_m$, such
that $\tau_{i}\sigma_{i}$ corresponds to the next line after $\tau'_i\sigma'_i$ according 
to the ordering defined in (iii). This will ensure that there are no fermion loops in the diagram. \\

(vi) Determine the symmetry factor of the graph, which is the number of
permutations of labeled lines and vertices that do not change its topology.\\

(vii) To calculate the contribution of the graph, integrate each {\em internal} 
time index between $0$ and $\beta$, sum over the {\em internal} spatial and spin 
indices, and divide the result by the symmetry factor. As an example, the contribution 
of the graph $c$ in Fig. \ref{fig:graph} is

\begin{eqnarray}
W(c)&=&\frac{1}{2}\sum_{1,2} (t_{j'1})^3(t_{12})^2t_{1j}\int_0^\beta d\tau_1\dots 
\int_0^\beta d\tau_6\nonumber\\
&\times&\sum_{\sigma_1\dots\sigma_6} C^0_2(\tau_1\sigma_1,\tau_3\sigma_3|\tau'\sigma,\tau_2\sigma_2)\nonumber\\
&\times&C^0_3(\tau_2\sigma_2,\tau_4\sigma_4,\tau_6\sigma_6|\tau_1\sigma_1,\tau_3\sigma_3,\tau_5\sigma_5)\nonumber\\
&\times&C^0_1(\tau_5\sigma_5|\tau_4\sigma_4)\ C^0_1(\tau\sigma|\tau_6\sigma_6).
\end{eqnarray}

(viii) To obtain the $l^{th}$ order contribution to the Green's function, add the 
contributions $W(c)$, of all the graphs with $l$ internal lines:

\begin{equation}
G^{(l)}=\sum_{c\ \in\ \textrm{order}\ l} W(c).
\end{equation}

In this scheme, the only zeroth order graph consists of a vertex and the two 
external lines. In the first order, the only possible topology has two vertices,
each having an external line, and a single internal line connecting them. 
In higher orders, the number of vertices can vary from two to $l+1$, where $l$ 
denotes the order, depending on the topology. The topologically distinct graphs 
up to the fourth order are shown in Fig. 4 of Ref.~\cite{w_metzner_91}.

\subsection{Computational implementation}

We have implemented a computer program to perform all of the steps described 
in Sec.~\ref{sec:steps} for the infinite-$U$ case. In this limit, since
no double occupancy is allowed, the calculation of the cumulants simplifies 
drastically. This enables us to carry out the expansion to eighth order.
In this subsection, we explain some of the details of this implementation at
each step.

\subsubsection{Generation of topologically distinct graphs}

To generate all topologically distinct diagrams in step (i) above, we need to have
a way of uniquely identifying them in a computer program. For this, we use the
concept of adjacency matrices. The elements of a $m\times m$ adjacency matrix,
where $m$ is the number of vertices, represent the connections between every two
vertices. For instance, for a graph with undirected lines between vertices, the
$(i,j)$ element can be an integer that simply counts the number of lines
between vertices $i$ and $j$. Here, since the lines are directed, we use a
generalization of this matrix where every element is replaced by an array of
size two. The first element of this array (we call it the left element) 
represents the number of incoming lines from vertex $i$ to vertex $j$ while 
the second element (or the right element) represents the number of outgoing 
lines from vertex $j$ to vertex $i$. 

One has to note that a topologically distinct graph cannot be uniquely 
represented by such an adjacency matrix since different labellings of the 
vertices, while not altering the topology, lead to different 
adjacency matrices. Therefore, one has to devise an algorithm to pick only
one, out of $m!$ possibilities, of the labellings of a graph to be able to
establish a one-to-one correspondence  between the graphs and its adjacency
matrix. This can be done, for example, through {\em sorting} the adjacency
matrix; by assigning the first row (column) to the vertex that possesses the 
largest number of lines, and so on. Alternatively, in our case, we can more
simply employ the order of vertices that results from rule (iii) above.

After defining the mapping between the adjacency matrices and graphs in the
computer algorithm, we generate graphs with $m$ vertices by considering all
possible numbers for the elements of the $m\times 2m$ adjacency matrix, subject
to the following two constraints: First, the number of incoming and
outgoing lines at each vertex have to be the same, so, if we subtract the sum of 
left elements and the
sum of right elements at each row (column) the result has to be zero. Second,
the total number of lines in the graph (or the sum of all elements of the
matrix, divided by 2) should be equal to the desired order in the expansion.
Note that, in this strong-coupling expansion, there is no line that leaves a
vertex and then enters the same vertex, i.e., the diagonal elements of all
adjacency matrices are zero.

\subsubsection{Cumulants}
\label{sec:cumulants}

We obtain cumulants to any order by taking functional derivatives of the generating 
functional with respect to Grassmann variables as described in Refs.~\cite{w_metzner_91,
OrlandNegele}. As a result, a cumulant of order $l$ is written in terms of the local unperturbed 
Green's function (UGF) of the same order, $G^0_l$, as well as lower order UGFs. 
In Appendix \ref{app:cmlnt}, we show this expansion for the first few cumulants. The 
calculation of the cumulants then reduces to the evaluation of the UGFs, which, 
for order $l$, is the expectation value of $2l$ time-ordered creation 
and annihilation operators with respect to the unperturbed Hamiltonian. 
For our case of the infinite-$U$ limit, since no double occupancy is 
allowed, a creation operator can only be followed by an annihilation operator 
and vice versa. Hence, the Green's function can 
assume only two distinct values depending on whether a creation or an 
annihilation operator is on the right side of the time-ordered product of operators.
The two values are, respectively, $(1-\rho)$ and $\frac{\rho}{2}$, where 
$\rho=\frac{2e^{\beta\mu}}{1+2e^{\beta\mu}}$ is the density in the atomic limit.
For example, we end up with the following terms for the first 
two orders:

\begin{widetext}
\begin{equation}
G^0_1(\tau_1\sigma_1|\tau'_1\sigma'_1)=\left<T_{\tau}c^{\dagger}_{j\sigma'_1}(\tau'_1)c_{j\sigma_1}(\tau_1)\right> 
=e^{\mu(\tau_1-\tau'_1)}\delta_{\sigma_1\sigma_1'}\left[\frac{\rho}{2}\Theta(\tau'_1-\tau_1)-(1-\rho)\Theta(\tau_1-\tau'_1) \right],\nonumber
\end{equation}
\begin{eqnarray}
G^0_{2}(\tau_1\sigma_1,\tau_2\sigma_2|\tau'_1\sigma'_1,\tau'_2\sigma'_2)&=&
\left<T_{\tau}c^{\dagger}_{j\sigma'_1}(\tau'_1)c_{j\sigma_1}(\tau_1)
c^{\dagger}_{j\sigma'_2}(\tau'_2)c_{j\sigma_2}(\tau_2)\right>\nonumber\\
&=&e^{\mu(\tau_1+\tau_2-\tau'_1-\tau'_2)}
\sum_{qp}(-1)^q(-1)^p\bigg[\frac{\rho}{2}\delta_{q\sigma_2p\sigma'_1}\delta_{q\sigma_1p\sigma'_2}
\Theta(p\tau'_1-q\tau_1)\Theta(q\tau_1-p\tau'_2)\Theta(p\tau'_2-q\tau_2)\nonumber\\
&+&(1-\rho)
\delta_{q\sigma_2p\sigma'_2}\delta_{q\sigma_1p\sigma'_1}
\Theta(q\tau_1-p\tau'_1)\Theta(p\tau'_1-q\tau_2)\Theta(q\tau_2-p\tau'_2)\bigg],
\end{eqnarray}
\end{widetext}
where the sum runs over permutations $p$ and $q$ of the time and spin indices of the 
primed and unprimed variables respectively, and $\Theta$ is the usual step function.

\subsubsection{Free multiplicities}

The spatial sums are performed for a specific lattice geometry. We have  
calculated them on the square lattice. In the computer program, we define a large 
enough lattice where we can fit any cluster with a number of sites at least 
twice as large as the maximum number of vertices in our largest order graphs.
We then assign vertices $j'$ (where an external line enters) and $j$ (where 
an external line exits) to two lattice sites with a given displacement between them.
The next part of the algorithm involves finding the number of possibilities for 
assigning the rest of vertices to lattice sites. This can be done by following the path 
we have defined for each graph in rule (iii) to go from vertex $j'$ to $j$. We start 
from vertex $j'$ and in each step, we move to the 
next vertex in the list and assign a site to it. We ensure that if we 
come back to a vertex in the graph, we also come back to the corresponding 
site on the lattice. However, since we are calculating free multiplicities, we
can assign the same lattice site to multiple vertices wherever the topology
of the graph allows for it. In Table.~\ref{tab:1}, we show the number of topologically 
distinct graphs in each order, along with the number of graphs that have nonzero
contributions on bipartite geometries, and the sum of free multiplicities for all graphs
in each order for the $(0,0)$ and $(1,0)$ separations, up to the 10th order.

This  computationally inexpensive process can be repeated for all possible separations 
(the maximum separation is set by the largest order considered). They can then be used to 
calculate the Fourier transform of the Green's function into the momentum space.

\begin{table}
\caption{Total number of topologically distinct graphs (second column), number of graphs that have
nonzero multiplicity on bipartite geometries (third column), and the sum of multiplicities of all graphs for
the smallest separations for which they have nonzero multiplicity (fourth column) at each order. The 
smallest separation for graphs with even number of lines (in even orders) is $r_{j'}-r_{j}=(0,0)$, and 
for graphs in odd orders is considered to be $r_{j'}-r_j=(1,0)$.}

\vspace{0.05in}

\begin{tabular}{|c|c|c|c|}
\hline
\color{blue} Order & \color{blue} Topo. Distinct &\color{blue} Used for Bipartite \color{blue}
&\color{blue} $\sum$ Multiplicities \\
\hline
\color{blue} 0 & 1 & 1 & 1 \\
\hline
\color{blue} 1 & 1 & 1 & 1 \\
\hline
\color{blue} 2 & 2 & 2 & 8 \\
\hline
\color{blue} 3 & 5 & 4 & 18 \\
\hline
\color{blue} 4 & 14 & 10 & 164 \\
\hline
\color{blue} 5 & 41 & 22 & 458 \\
\hline
\color{blue} 6 & 130 & 59 & 4240 \\
\hline
\color{blue} 7 & 431 & 146 & 13544 \\
\hline
\color{blue} 8 & 1512 & 425 & 130516 \\
\hline
\color{blue} 9 & 5542 & 1136 & 448211 \\
\hline
\color{blue} 10 & 21236 & 3497 & 4408216 \\
\hline
\end{tabular}
\label{tab:1}
\end{table}

\subsubsection{Time integrals}

As seen in Sec.~\ref{sec:cumulants}, the cumulants for the infinite-$U$ Hubbard 
model consist of products of only step functions and exponentials in the internal and 
external imaginary times. After multiplying several cumulants to obtain the contribution 
of a graph, we typically end up with a huge number of terms, each consisting of the 
product of a set of step functions of the time variables, the exponentials associated 
with the external times (the exponentials associated with the internal times cancel), 
Kronecker delta functions of the spin indices, and a function of $\rho$. 
As mentioned before, one of the main advantages of 
our approach is that the time integrals over internal time variables can be 
taken independently of the spatial sums (free multiplicity calculations). We 
choose $\tau'=0$ without loss of generality since the Green's function is
a function of $\tau-\tau'$ and $G(\tau-\tau'<0)$ can be obtained from $G(\tau-\tau'>0)$
using the anti-periodicity of the Green's function in imaginary time~\cite{agd}. To see 
how the time integrals are evaluated, we proceed with the following example. Suppose 
that one of the terms that belongs to a graph in the third order can be written as:
\begin{equation}
\mathcal{I}(\tau)=\int_0^{\beta}\int_0^{\beta}\int_0^{\beta}d\tau_1d\tau_2  d\tau_3 
\Theta(\tau_1-\tau_3)\Theta(\tau-\tau_3). \label{eq:I} 
\end{equation}
Note that in the above example, we have a smaller number of step functions
in the integrand than typically expected for a term in the third order. However, 
the above combination is a perfectly valid one as the products of step functions 
are often simplified given that $\Theta^n(x)=\Theta(x)$ for any nonzero $n$.
The integral over $\tau_2$ yields a factor $\beta$ as there is no restriction 
on $\tau_2$. The remaining integrals are nonzero if $\tau_1>\tau_3$ and 
$\tau>\tau_3$. But, the latter condition does not uniquely determine the 
position of $\tau_1$ relative to $\tau$ in the $[0,\beta]$ interval.
Therefore, we consider the two possibilities, $\tau>\tau_1$ and $\tau<\tau_1$, and
rewrite the integral of Eq.~(\ref{eq:I}) as 
\begin{eqnarray}
\label{eq:I2}
&\mathcal{I}(\tau)&=\beta\int_0^{\beta}\int_0^{\beta}d\tau_1d\tau_3 \\
&\times&\left[\Theta(\tau_1-\tau)\Theta(\tau-\tau_3) +
\Theta(\tau-\tau_1)\Theta(\tau_1-\tau_3) \right].\nonumber
\end{eqnarray}
Note that for any value of $\tau_1$ and $\tau_3$, only one of the terms in the 
integral in
Eq.~(\ref{eq:I2}) is nonzero, justifying the equality. At this point, we can use
the known results for the types of integrals in Eq.~(\ref{eq:I2}) (see Appendix 
\ref{app:TI}), leading to $\beta[\tau(\beta-\tau)+\frac{\tau^2}{2!}]$.

Computationally, the two distinct possibilities for the ordering of times in 
the above example can be found by generating all of the permutations of the time
indices, and for each permutation, examining whether every step function 
in the product is nonzero. If that is the case, a multiplication of step functions
corresponding to that permutation is inserted as the integrand.

\subsubsection{Symmetry factor}

Calculating the symmetry factor of each graph is straightforward in the 
framework of adjacency matrices. First, we note that the symmetry factor 
is proportional to the factorials of elements of the adjacency matrix in 
its upper triangle as they correspond to the number of permutations of 
directed lines that do not change the topology 
of the graph. Second, in order to find the symmetry factor related to those 
permutations of labeled vertices that leave the graph topology 
intact, we simply generate all the $m!$ matrices that correspond to different 
orderings of vertex labels and find how many of them are the same as the 
original matrix. We then multiply this number by the factorials calculated 
in the first step to obtain the symmetry factor of the graph.

\section{Analytical Results\label{GreenSigma}}

After evaluating the contribution of each diagram in a particular order 
by multiplying its free multiplicity for a given separation, time 
integral, and the spin sum, and dividing it by the symmetry factor,
we add all of those contributions for that order to form the Green's 
function in terms of the atomic density, $\rho$, the imaginary time
$\tau$, $t$, $\mu$, and $\beta$. By calculating the spatial sums for 
all possible separations for each graph and
performing a Fourier transformation on the space and imaginary time, 
one can express the Green's function in terms of the momentum, $k$,
and the Matsubara frequency, $\omega_n$. Below, we show the resulting
Green's function in the first four orders \cite{compare}:

\begin{widetext}

\begin{eqnarray}
G^{(0)}_{\sigma}(z,k)&=&\frac{1-\frac{\rho }{2}}{z}, \nonumber \\
G^{(1)}_{\sigma}(z,k)&=&\frac{\left(1-\frac{\rho }{2}\right)^2 \epsilon _k}{z^2},\nonumber \\
G^{(2)}_{\sigma}(z,k)&=&\frac{\left(1-\frac{\rho }{2}\right)^3 \epsilon _k^2}{z^3}+\frac{(4-\rho ) \rho 
   \left(1-\frac{\rho }{2}\right) t^2}{z^3}-\frac{2 \beta  (\rho -1) \rho 
   t^2}{z^2}+\frac{\beta ^2 \rho  [(3-2 \rho ) \rho -1] t^2}{z},\\
G^{(3)}_{\sigma}(z,k)&=&  \frac{(1-\frac{\rho}{2})^4 \epsilon _k^3}{z^4}-\frac{7 (\rho -4) \rho  (\rho -2)^2 t^2\epsilon _k}{16 z^4}+\frac{3  \beta  (\rho -1) \rho  (\rho -2) t^2 \epsilon _k}{2 z^3}+\frac{\beta ^2 (\rho -1) \rho  [\rho  (7 \rho
   -19)+8] t^2 \epsilon _k}{4 z^2}, \nonumber\\
G^{(4)}_{\sigma}(z,k)&=&  \frac{(1-\frac{\rho}{2})^5 \epsilon _k^4}{z^5}+\frac{5 (\rho -4) \rho  (\rho -2)^3 t^2\epsilon _k^2}{16  z^5}-\frac{\rho  \{\rho  [(\rho -8) \rho -152]+240\} (\rho
   -2)t^4}{8 z^5} \nonumber\\
   &-&\frac{\beta  (\rho -1) \rho  (\rho -2)^2 t^2\epsilon _k^2}{z^4} +\frac{\beta  (\rho -1) \rho  [\rho  (4 \rho +11)-16]t^4}{z^4} -\frac{\beta ^2 (\rho -1) \rho  [\rho  (5 \rho -14)+6] (\rho -2) t^2 \epsilon _k^2}{4 z^3} \nonumber \\
   &+&\frac{\beta ^2 (\rho -1) \rho \{\rho  [2 \rho  (5 \rho -24)+43]-16\}t^4}{2 z^3} 
   -\frac{\beta ^3 (\rho -1) \rho  [\rho  (97 \rho -100)+18]t^4}{6  z^2} \nonumber \\
   &-&\frac{\beta ^4 (\rho -1) \rho  \{\rho  [\rho  (388 \rho -591)+236]-18 \} t^4}{24 z} \nonumber\\
      &\vdots&, \nonumber
\end{eqnarray}

\end{widetext}
where $z=i\omega_n+\mu$, and $\epsilon_k=-2t[\cos(k_x)+\cos(k_y)]$. Note that in this format, 
the Green's function is written in terms of the atomic density $\rho$ or 
equivalently the chemical potential $\mu$,  and not the true 
density for the many-body system, $n=1+G_{jj\sigma}(\tau-\tau'=0^+,\mu)$ \cite{note2}. 
By definition, $n$, too, can be written as an expansion in the hopping (using the expansion
for the local Green's function). However, we would like to treat $n$ as a parameter and re-write
the Green's function in terms of it. In that 
case, the chemical potential can no longer remain constant and we have to 
solve for it order by order in terms of $n$ and $t$: $\mu=\mu^{(0)}+\mu^{(2)}
+\mu^{(4)}\dots$ where
\begin{widetext}
\begin{eqnarray}
 n-1 &=& G^{(0)}_{jj\sigma}(0^+, \mu^{(0)}) +G^{(2)}_{jj\sigma}(0^+, \mu^{(0)})+\frac{dG^{(0)}_{jj\sigma}(0^+,\mu)}{d\mu}\vert_{\mu=\mu^{(0)}}\mu^{(2)} + G^{(4)}_{jj\sigma}(0^+, \mu^{(0)})\nonumber\\
&&+\frac{dG^{(2)}_{jj\sigma}(0^+,\mu)}{d\mu}\vert_{\mu=\mu^{(0)}}\mu^{(2)}+\frac{1}{2}\frac{d^2G^{(0)}_{jj\sigma}(0^+,\mu)}{d\mu^2}\vert_{\mu=\mu^{(0)}}(\mu^{(2)})^2+\frac{dG^{(0)}_{jj\sigma}(0^+,\mu)}{d\mu}\vert_{\mu=\mu^{(0)}}\mu^{(4)}  + \ldots
\label{eq:xdef}
\end{eqnarray}
\end{widetext}
Inverting  this equation for $\mu$ in terms of  $n$, we obtain
\begin{eqnarray}
\mu^{(0)}&=& \frac{1}{\beta}\log\frac{n}{2(1-n)}, \nonumber \\
\mu^{(2)}&=&2(2n-1)t^2\beta, \nonumber \\
\mu^{(4)}&=&\frac{1}{12}(6+n(n-4)(1+4n))t^4\beta^3 \\
&\vdots& \nonumber
\end{eqnarray}

Finally, by inserting these back into the expansion for the momentum- and 
frequency-dependent Green's function order by order, we end up with the 
following terms for up to the fourth order~\cite{note1}:

\begin{widetext}

\begin{eqnarray}
G^{(0)}_{\sigma}(z,k)&=&\frac{1-\frac{n}{2}}{z}, \nonumber \\
G^{(1)}_{\sigma}(z,k)&=&\frac{(1-\frac{n}{2})^2 \epsilon _k}{z^2}, \nonumber \\
G^{(2)}_{\sigma}(z,k)&=&\frac{(1-\frac{n}{2})^3 \epsilon _k^2}{z^3}+\frac{[2 (n-2)-n] [2 (n-1)-n] n t^2}{2 z^3}-\frac{[2 (n-1)+n] t^2 \beta }{z^2}, \nonumber \\
G^{(3)}_{\sigma}(z,k)&=&\frac{(1-\frac{n}{2})^4 \epsilon _k^3}{z^4}-\frac{7 [2 (n-2)-n] n (2-n)^2 t^2 \epsilon _k}{16 z^4}
-\frac{[2 (n-1)-n] [2(2-3 n)+(n-1) n] t^2 \beta  \epsilon _k}{2 z^3}\nonumber\\
&-&\frac{(n-1)^2 n^2 t^2 \beta ^2 \epsilon _k}{4 z^2},\nonumber \\
G^{(4)}_{\sigma}(z,k)&=&+\frac{(1-\frac{n}{2})^5 \epsilon _k^4}{z^5}+\frac{5 [2 (n-2)-n] [2 (n-1)-n]^3 n t^2 \epsilon _k^2}{16 z^5}\nonumber \\
&+&\frac{[2 (n-1)-n] n \left(-n^3+8 n^2+152 n-240\right) t^4}{8 z^5}+\frac{\{4 (n-1) n-6 [2 (n-1)+n]\} (2-n)^2 t^2 \beta  \epsilon _k^2}{8 z^4}\nonumber \\
   &+&\frac{n\{2 [-3 n^2+6 (2 n-3) n+4 (n-1) (9 n-10)]+(1-n) n (4 n+2)\} t^4 \beta }{2 z^4}\nonumber\\
   &+&\frac{[4 (2 n^3+6 n^2-10 n+4)-(n-1) n^2 \left(4 n^2-12 n+2\right)] t^4 \beta ^2}{4 z^3}+\frac{\left(-6 n^3+68 n^2-20 n-24\right) t^4 \beta ^3}{48 z^2}\nonumber \\
   &\vdots&, \label{Gn}
\end{eqnarray}

\end{widetext}
where $z= i\omega_n + \mu^{(0)}$. It is perhaps even more useful to extract a self-energy from this expansion.  
The Dyson-Mori self-energy ( denoted simply with $\Sigma_{DM}\to \Sigma$)  can  be deduced using $\Sigma(z,k)=z-a_{\scriptscriptstyle G}[\epsilon_k+G(z,k)^{-1}]$, where $a_{\scriptscriptstyle G}=(1-n/2)$
\cite{s_shastry_11}:

\begin{widetext}

\begin{eqnarray}
\Sigma^{(0)}(z,k)&=&0,\nonumber\\
\Sigma^{(1)}(z,k)&=&0,\nonumber\\
\Sigma^{(2)}(z,k)&=&\frac{t^2 \beta(6 n-4 )}{n-2}-\frac{\left(n^2-4 n\right) t^2}{z},\nonumber\\
\Sigma^{(3)}(z,k)&=&\frac{n^2 t^2 \epsilon _k\beta^2 \left(1+ n^2-2 n\right)}{2 (n-2)}-\frac{(n-4) (n-2) n t^2
   \epsilon _k}{8 z^2}+\frac{n t^2 \epsilon _k \beta( n-1 )}{z},\nonumber \\
\Sigma^{(4)}(z,k)&=&\frac{t^4 \beta^3 \left(12 +3 n^3-34 n^2+10 n\right)}{12 (n-2)}-\frac{3
   \left(n^4-8 n^3+72 n^2-80 n\right) t^4}{4 z^3}\nonumber\\
   &+&\frac{2 t^4\beta \left(2 n^4-40  n^3+65 
   n^2-24  n\right)}{(n-2) z^2}+\frac{t^4 \beta^2 \left(2 n^6-12  n^5+19  n^4-19
    n^3+10  n^2\right)}{(n-2)^2 z} \nonumber \\
      &\vdots&
      \label{Sigma}
\end{eqnarray}

\end{widetext}

\section{Convergence and the Pad\'e Approximation\label{Pade}}

\begin{figure}[t]
\centerline {\includegraphics*[width=3.3in]{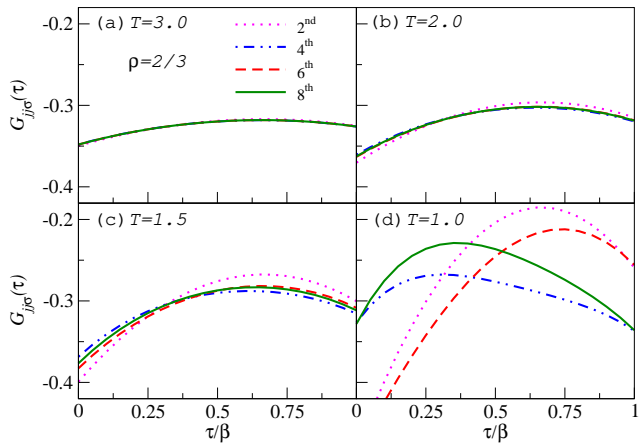}}
\caption{Local Green's function for a constant chemical potential $\mu=0$ vs imaginary time
at (a) $T=3.0$, (b) $T=2.0$, (c) $T=1.5$, and (d) $T=1.0$. $t=1$ is the unit of energy throughout 
the paper.}
\label{fig:Gloc}
\end{figure}

In Fig.~\ref{fig:Gloc}, we show the local imaginary time Green's 
function for $\rho=2/3$, corresponding to $\mu=0$, at different
temperatures. At $T=3.0$ (unless specified otherwise, we take $t=1$ as the unit of energy and work 
in units where $k_B=1$ throughout the paper), the series show very good 
convergence as expected in this high temperature region. Note that 
odd terms in the series are zero for this local quantity. As we lower 
$T$ to 2.0, there are some discrepancies between low orders, but the
last two orders (6 and 8) still agree very well. This is no longer 
the case as we get closer to $T=1$, below which the finite series is 
divergent by definition. This is because in the absence of any other 
energy scale in the system, an expansion in $t$ can be viewed as an
expansion in $\beta$. In other words, $\beta^{m+1}$ always couples to $t^m$
in the series for the Green's function.
In Fig.~\ref{fig:Gloc}(d), one can see large fluctuations between 
different orders already at $T=1.0$ and there is no clear picture
from the bare results as to what the actual shape of the Green's function is.

\begin{figure}[t]
\centerline {\includegraphics*[width=3.3in]{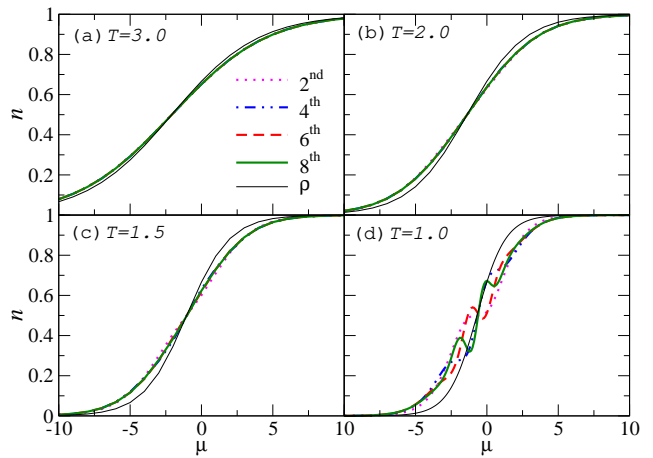}}
\caption{Density $n$ as a function of the chemical potential 
at (a) $T=3.0$, (b) $T=2.0$, (c) $T=1.5$, and (d) $T=1.0$. Thin solid
lines are the density in the atomic limit, $\rho=\frac{2e^{\beta\mu}}{1+2e^{\beta\mu}}$.}
\label{fig:dens_mu}
\end{figure}

To demonstrate the trends in the convergence of the series at other values
of $\mu$, in Fig.~\ref{fig:dens_mu}, we show the equation of state at 
the same four temperatures as in Fig.~\ref{fig:Gloc}. We also show the 
equation of state in the atomic limit ($\rho$ vs $\mu$). We find that 
the last two orders more or less agree with each other for all 
$\mu$ at $T\gtrsim1.5$. However, for $T=1$, the convergence is lost 
in the vicinity of $\mu=0$. This shows that the poor convergence of the 
local Green's function at this value of $\mu$, seen in Fig.~\ref{fig:Gloc}(d),
represents the worst case scenario. An important feature of the equation 
of state as observed in Fig.~\ref{fig:dens_mu} is that even at these high 
temperatures, there are significant deviations of the many-body density from 
the density in the atomic limit near the extreme limits of $n=0$ and $n=1$.

\begin{figure}[t]
\centerline {\includegraphics*[width=3.3in]{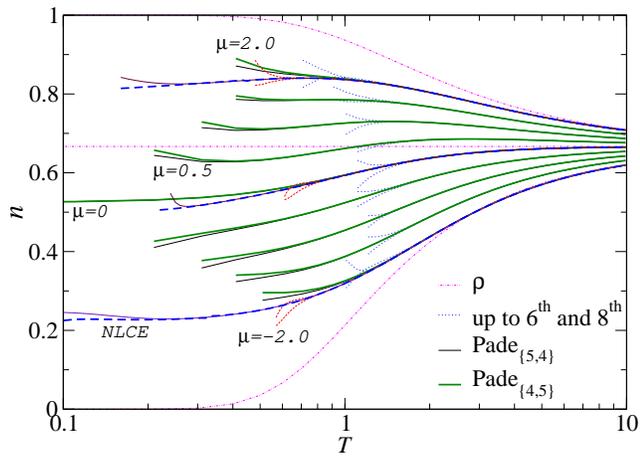}}
\caption{Average density $n$ as a function of temperature for a range of $\mu$ 
from -2.0 to 2.0, with the increment of $0.5$. The two indices of \Pd in curly 
brackets indicate the order of the polynomials in the numerator and the denominator. 
From bottom to top, the dotted-dashed magenta lines are $\rho$ for
$\mu=-2$, 0, and 2. We are also showing results from the NLCE for these
three values of the chemical potential as thin dashed red lines (last two orders 
of the bare sums), and thick dashed blue and thin solid violet lines (after 
Wynn resummations with five and four cycles of improvement, respectively)~\cite{NLCE-tJ}.}
\label{fig:dens_T}
\end{figure}

It is instructive now to study the temperature dependence of the density 
at a given $\mu$, and to find out how the region of convergence can be 
extended in temperature by the use of Pad\'{e} approximations. In Fig.~\ref{fig:dens_T}, 
we show the temperature dependence of the density for various positive and 
negative values of $\mu$. We show the direct sums as well as results after 
two different \Pd approximations. The results in the atomic limit $[\rho(T)$] 
are shown for $\mu=0$ and $\pm 2.0$. In the atomic limit, 
the system has two ground-states depending on the sign of $\mu$. They 
correspond to $\rho=1$ and  $\rho=0$ for positive and negative $\mu$, respectively. 
At exactly $\mu=0$, $\rho$ is temperature-independent at $2/3$. As one can see 
in Fig.~\ref{fig:dens_T}, the real density for the many-body system has a 
qualitatively different behavior than $\rho$ starting at relatively high temperatures.
The temperature where $n$ starts deviating from $\rho$ due to correlations 
is around $T\sim2$ for $\mu=-2$ and $T\sim5$ for $\mu=2$. As expected, the 
density for $\mu=0$ falls below $2/3$ for all $T$. To perform \Pd approximation
for $n$ vs $T$, we first expand $\rho$, i.e.  the zeroth order term, in powers
of $\beta$ and then add the rest of the higher order terms from the series. 
Therefore, in the case of $\mu=0$, where $\rho$ is temperature independent, 
the odd powers of $\beta$ in the series for $n$ vanish and the two \Pd approximants yield 
the same function, leading to $n\sim0.525$ for the ground state. Nevertheless, we
cannot verify that this is the true value of the ground-state density of the system 
for $\mu=0$. 

The static properties of the model, such as the density, can in principle be 
obtained in higher orders by avoiding the relatively difficult calculation of 
the Green's function, and calculating only the free energy instead. However,
for this purpose, we can also take advantage of the novel NLCE 
method that has been developed in recent years~\cite{NLCE-tJ}. 
NLCE uses the same basis as high-temperature expansions, but calculates 
properties of finite clusters exactly, as opposed to perturbatively, using
full diagonalization techniques. As a result, the convergence region of the 
NLCE is typically extended to lower temperatures in comparison to
high-temperature expansions with the same number of terms.

In Fig.~\ref{fig:dens_T}, we show results from the NLCE for the \tj model with $J=0$
for up to the 11th order in the site expansion,  where contributions of all clusters 
with up to 11 sites are considered, for $\mu=0$ and $\pm 2.0$.
By comparing the direct sums in NLCE (thin dashed red lines represent the last
two orders) with those from our series, we find that while we have perfect
agreement between NLCE and the converged bare sums in the series,
the \Pd approximants overestimate the value
of $n$ in all cases at temperatures lower than one.
The convergence of the NLCE results at low temperatures can be further improved 
using numerical resummations. Here, we show those obtained from the Wynn
algorithm~\cite{NLCE-tJ} by thin solid violet and thick dashed blue lines.
Remarkably, the convergence is extended to $T\sim0.2$ for $\mu=-2.0$, and
$T\sim0.3$ for $\mu=0$ and 2.0. The results for $\mu=0$ show that the ground
state density is likely less than $0.525$.

\begin{figure}[t]
\centerline {\includegraphics*[width=3.3in]{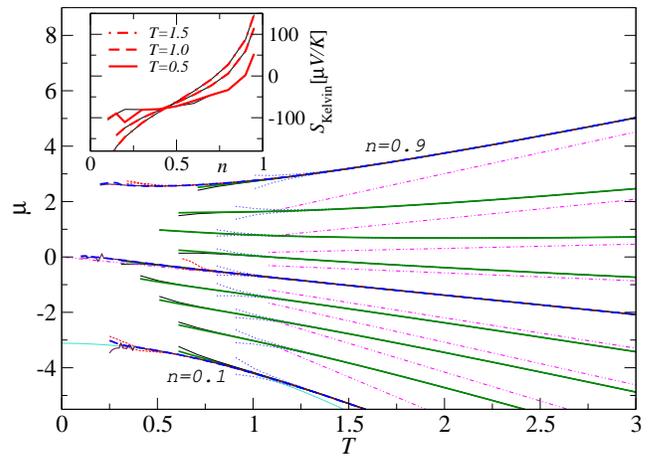}}
\caption{Chemical potential $\mu$ at fixed density vs temperature for densities 
from $n=0.1$ to $n=0.9$ (from bottom to top with the increment of $\Delta n=0.1$). 
The lines are the same as in Fig.~\ref{fig:dens_T}, except that the dotted-dashed 
magenta lines are the zeroth order of the chemical potential in the atomic limit, 
i.~e., $\mu^{(0)}=T\log\frac{n}{2(1-n)}$, and that thin solid lines are Pad{\'e}$_{\{6,3\}}$. 
Here, we show the NLCE results for $n=0.1$, 0.5, and 0.9. The light blue (light gray) 
solid line is the fit of the low-temperature NLCE results for $n=0.1$ after resummation 
to $A+BT^2$ with $A=-3.12$ and $B=-1.10$. 
The inset shows the Kelvin thermopower,  $S_{\textrm{Kelvin}}$, from NLCE as 
defined in Eq.~(\ref{eq:kelvin}), in units of microvolts per degree Kelvin vs density.
At each temperature, the two lines correspond to different Wynn resummations.}
\label{fig:mu_T}
\end{figure}

In Fig.~\ref{fig:mu_T}, we plot the chemical potential of the system as a function 
of temperature for various fixed densities by inverting functions such as those 
seen in Fig.~\ref{fig:dens_T}. Here, the dotted dashed lines represent the zeroth
order chemical potential $\mu^{(0)}$ for a fixed density. They all approach zero as
$T\to0$ since they correspond to the atomic limit. The results from the series and 
the NLCE suggest a different behavior starting at relatively high temperatures for the
correlated system, except for the density near 0.5, where the linearity of the chemical 
potential, and the coincidence with the results from the atomic limit, is extended to low 
temperatures. This is consistent with the $\mu=0$ curve in Fig.~\ref{fig:dens_mu} approaching 
$n\sim0.5$ at low temperatures. On the other hand, in the low density Fermi liquid 
regime, the low-temperature chemical potential is expected to be proportional to 
$T^2$. We find that the resumed NLCE results
for $n=0.1$ agree with this behavior as they provide a reasonable fit to the function 
$A+BT^2$, as shown by a light blue (light gray) line in Fig.~\ref{fig:mu_T}.

Another feature seen in the plots of chemical potential at fixed density, with potentially 
important implications for the state of the system, is the change in sign of the slope of $\mu$ 
vs $T$ at low temperatures. Recent theories of thermopower of correlated systems 
identify the Kelvin formula for thermopower \cite{p-s,stp} by the expression
\begin{equation}
S_{\textrm{Kelvin}}= \frac{-1}{q_e} \left(\frac{\partial \mu}{\partial T}\right)_{N,V}=  
\frac{1}{q_e} \left(\frac{\partial S}{\partial N}\right)_{T,V},
\label{eq:kelvin}
\end{equation}
where $q_e= -|e|$ is the electron charge, $S$ the entropy and a Maxwell relation 
is employed in the second identity.  This formula captures the considerations of  Kelvin's 
famous paper on reciprocity in 1854 \cite{kelvin}, within  a contemporary setting. 
As explained in Refs. \cite{p-s,stp}, this expression represents the ``thermodynamic'' contribution 
to the true thermopower in addition to the dynamical contributions, that are assumed small 
in many correlated systems and neglected here. We see from this expression  that a flat 
chemical potential in temperature implies a maximum in entropy at the corresponding density, 
and locates a density where the thermopower changes sign (from electronlike to hole like), 
as often seen in correlated systems. From Fig.~\ref{fig:mu_T}, we observe that 
$\frac{\partial \mu}{\partial T}>0$ and hence the Kelvin  thermopower is positive for densities 
close to half filling, whereas near the empty band things are reversed and we get electron 
like thermopower. The change in sign seems to arise at a density $n$ between $0.7$ and 
$0.9$, somewhat greater than the value $n=\frac{2}{3}$ from the naive atomic limit. A detailed 
discussion of the thermopower, and the related Hall constant  in cuprates and in the two-dimensional \tj model can be found in 
Refs. \cite{stp,Garg}.

\begin{figure}[t]
\centerline {\includegraphics*[width=3.3in]{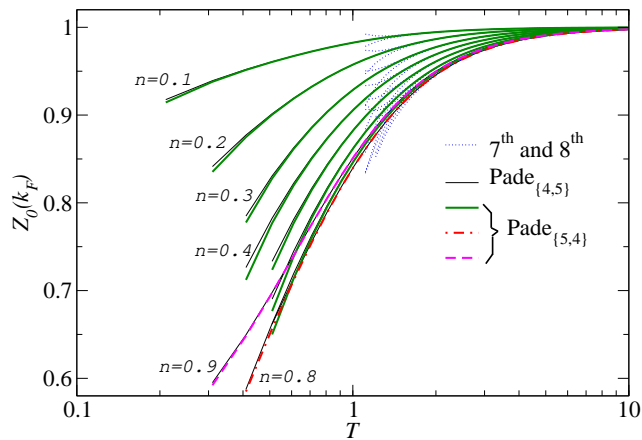}}
\caption{The quasiparticle fraction, defined in the Matsubara frequency space, 
at the nodal Fermi surface of the corresponding free Fermi gas, Eq.~(\ref{eq:qp}), 
after Pad\'{e} approximation vs temperature for different values of density. At 
temperatures below one, the quasiparticle fraction initially decreases with increasing 
the density before increasing again for $n>0.7$. The green thick solid lines are for 
$n=0.1\dots 0.7$ from top to bottom.}
\label{fig:qp_T}
\end{figure}

In Fig.~\ref{fig:qp_T}, we show the analog of the quasiparticle fraction defined in 
the Matsubara frequency space as
\begin{equation}
\label{eq:qp}
Z_0(k)=\left[1-\frac{{\rm Im} \Sigma(\omega_0,k)}{\omega_0}\right]^{-1},
\end{equation}
where $\omega_0=\pi T$ is the lowest Matsubara frequency, as a function of temperature
at various densities. We choose the momentum $k$ to be the nodal Fermi vector of 
a free Fermi gas with the same density ($k_F$). Previous studies based on the 
ECFL~\cite{ecfl}, or high-temperature expansions~\cite{hte}, suggest that this model 
possesses a Fermi surface coinciding with that of the free Fermi gas. The 
quantity in Eq.~(\ref{eq:qp}) will be equal to the actual quasiparticle fraction deduced 
from the self-energy in the real frequency axis,
$Z(k)=[1-\frac{\partial \Sigma(\omega,k)}{\partial\omega}|_{\omega\to 0}]^{-1}$, in 
the limit $T\to0$. Therefore, the lowest temperatures we have access to may not
be low enough to provide us with useful insight as to how the ground-state 
value of this quantity may vary with density. However, already at $T\sim0.5$, 
\Pd approximants offer an unexpected insight. We find that $Z_0(k)$
decreases monotonically by increasing the density for $n<0.8$, then increases
as $n$ increases to $0.9$. Interestingly, the onset of this change of behavior coincides 
with that of the change of sign in the thermopower discussed earlier. As $n\to 1$, we do expect the true ground-state  value of $Z(k)$ to vanish, therefore this non monotonic dependence is presumably an  artifact resulting from  the finite T definition employed.

\begin{figure}[t]
\centerline {\includegraphics*[width=3.3in]{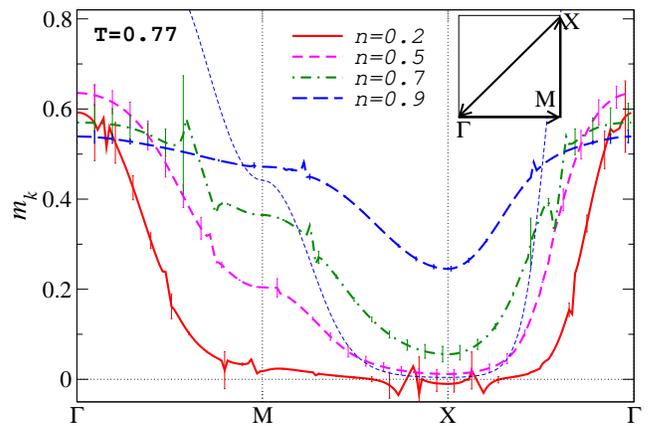}}
\caption{Momentum distribution function at $T=0.77$ for $n=0.2$, 0.5, 0.7, and 0.9 vs 
momentum, as obtained from the average of the two \Pd approximations ($\{4,5\}$ and 
$\{5,4\}$), around the irreducible wedge of the Brillouin zone as shown in the inset. 
Vertical lines show the difference between the two \Pd approximants. The thin
dashed line is the momentum occupation number of a free Fermi gas for $n=0.9$ at
the same temperature.}
\label{fig:mom_dist}
\end{figure}

In Fig.~\ref{fig:mom_dist}, we show the momentum occupation number, 
$m_k=\langle c^{\dagger}_{k\sigma}c_{k\sigma}\rangle$, versus $k$ at $T=0.77$ for different 
total densities. Features of this quantity at much lower temperatures were discussed in 
Ref.~\cite{hansen-shastry} for the \tj model. However, the value of the density in the latter 
study was limited to $n\lesssim0.75$. Here, we find that even at high temperatures, as the 
density approaches half filling, there is a huge redistribution of occupations in comparison 
to the free Fermi gas, as evidenced by the difference in $m_k$ for $n=0.9$ between the two cases
as seen in Fig.~\ref{fig:mom_dist}.

\begin{figure}[b]
\centerline {\includegraphics*[width=3.3in]{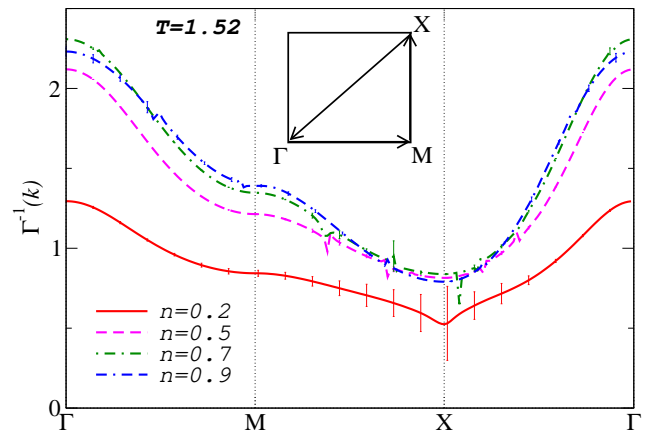}}
\caption{Inverse lifetime, defined in Eq.~(\ref{eq:life}), at $T=1.52$ and for $n=0.2$, 0.5, 
0.7, and 0.9 vs momentum around the irreducible wedge of the Brillouin zone shown 
in the inset. $t=1$ sets the unit of energy. Lines are the same as in Fig.~\ref{fig:mom_dist}.}
\label{fig:lifetime_T}
\end{figure}

In a recent publication~\cite{e_khatami_13}, the first moments of the electronic spectral 
functions of this model were studied using the same series expansion. It was shown that 
a modified first moment,   
(the ``greater" moment) can better capture the location of the spectral peak at higher densities 
than the symmetric first moment. More information about the spectral properties of electrons in this 
model can be gathered from higher order moments, also accessible through the series. 
In Fig.~\ref{fig:lifetime_T}, we show the width of the quasiparticle peak, or the inverse lifetime, 
defined as
\begin{equation}
\label{eq:life}
\Gamma^{-1}(k)=\sqrt{\varepsilon^{>}_2(k)-[\varepsilon^{>}_1(k)]^2},
\end{equation}
where $\varepsilon^{>}_1(k)$ and $\varepsilon^{>}_2(k)$ are the first and second greater 
moments, respectively, obtained from the series as described in Eq.~(7) of Ref.~\cite{e_khatami_13}.
Since the spectral function is largely skewed at higher densities \cite{Asymm}, the width generally 
grows as the density increases.

\section{Spectral Functions}

We next turn to a study of the spectral functions $\rho_G(k,\omega)$, denoted by $A(\omega,k)$ 
in standard photoemission studies. This can be found from the usual relation
$\rho_G(\omega,k)\equiv -\frac{1}{\pi} {\rm Im} \  G(\omega + \mu^{(0)}+i\eta,k)$, and requires 
a knowledge of the Greens function for complex frequencies.
To extract spectral functions, we represent our Green's function as a continued fraction, which, 
when Taylor expanded to eighth order, reproduces \disp{Gn}. That is, we write $G$ as (see 
Ref.~\cite{moments} for the notation) 
\begin{eqnarray}
\hspace{-9mm} \text{$G(z,k) = \frac{a_{\scriptscriptstyle G}}{z+b_1-}\;\; \frac{a_1}{z+b_2-}\;\; 
\frac{a_2}{z+b_3-} \;\;\frac{a_3}{z+b_4-} \;\; \frac{a_4}{z},$} \nonumber\\ 
\label{cfraction}\end{eqnarray}
where $a_l >0$ and $b_l$ are real.  As explained in \refdisp{moments} (see also \cite{Pairault}), 
these conditions ensure that the resulting spectral function obtained from analytic continuation 
is positive definite. The formulas for the $a_l$ and $b_l$ can be obtained by suitably combining 
the ``raw'' moments; this procedure is detailed  in \refdisp{Dupuis}. In the 
infinite-$U$ Hubbard model, we know {\em a priori} how many floors will be in the continued fraction representation 
of a Green's function series of a given order. This is because the constants $b_l$ have units of energy (and 
must therefore to leading order go like $t$), and the constants $a_l$ have units of energy squared (and must 
therefore to leading order go like $t^2$). Therefore, we know that \disp{cfraction} is the correct, i.e., maximal 
continued fraction form obtainable  from our eighth-order series. This is  an advantage over the case of the 
finite-$U$ Hubbard model [see \refdisp{Pairault}], where the presence of the energy scale $U$ means that 
the number of floors necessary to represent a series of a given order must be determined empirically.

\begin{figure}[t]
\includegraphics[width=\columnwidth]{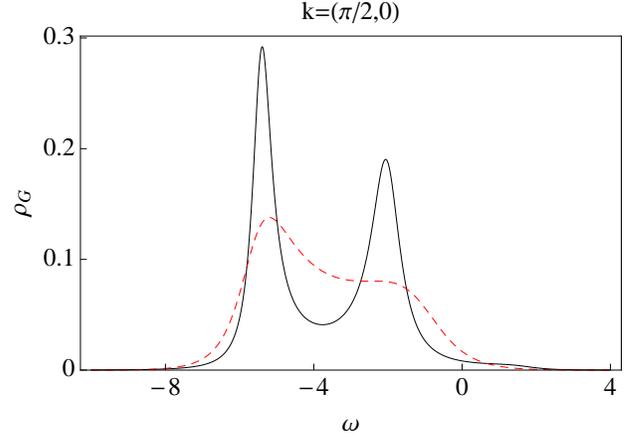}
\caption{The spectral density for the physical Green's function vs
$\omega$ for $T=1.1$ and $n=0.9$. $t=1$ sets the unit of energy.  
The red (dashed) curve is obtained from the 
TM scheme with the self-energy  \disp{rhoSigGausssymm} and the black (solid)
curve is obtained from the TM scheme with the second level self-energy 
[\disp{rhoSig2Gauss}]. The latter accentuates the unphysical secondary peak 
of the TM scheme spectral function. }
\label{compare02}
\end{figure}

\begin{figure*}[h]
\begin{center}
\includegraphics[width=.68 \columnwidth] {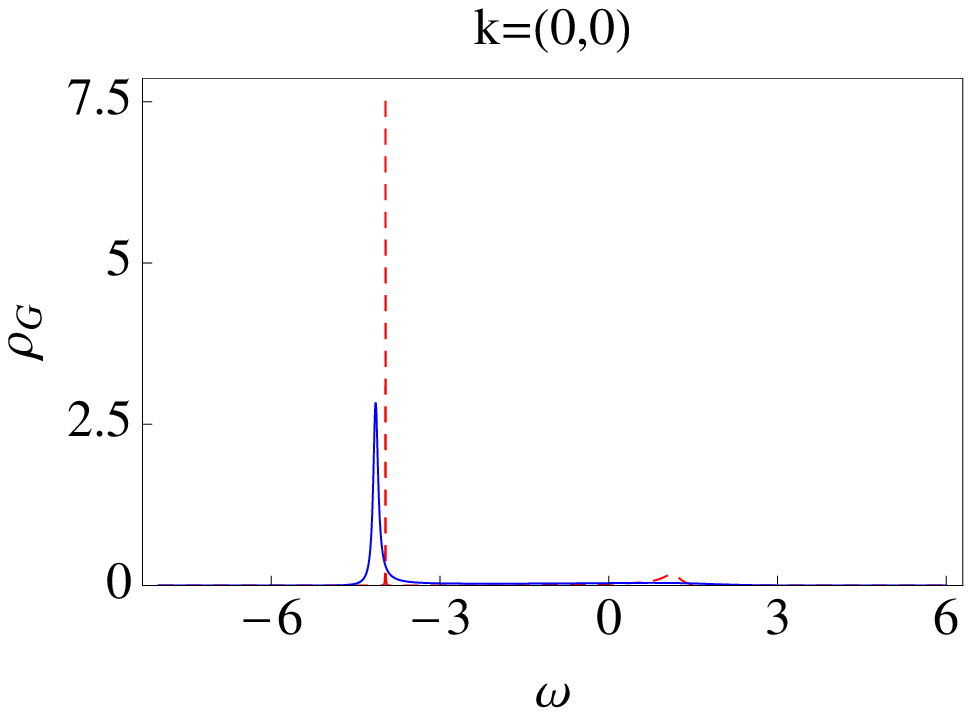}
\includegraphics[width=.68 \columnwidth] {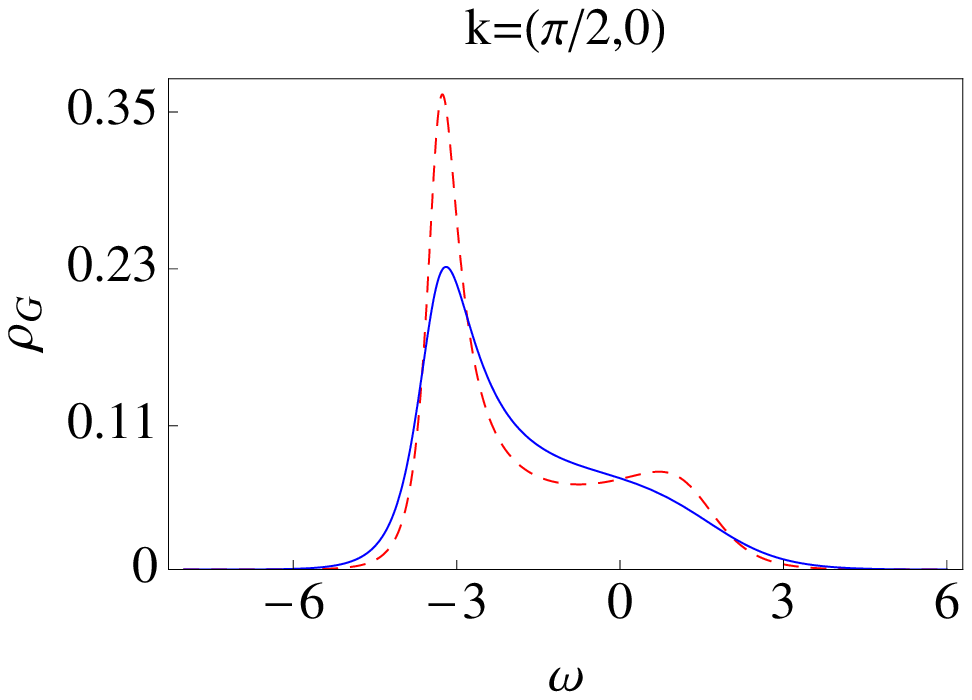}
\includegraphics[width=.68 \columnwidth] {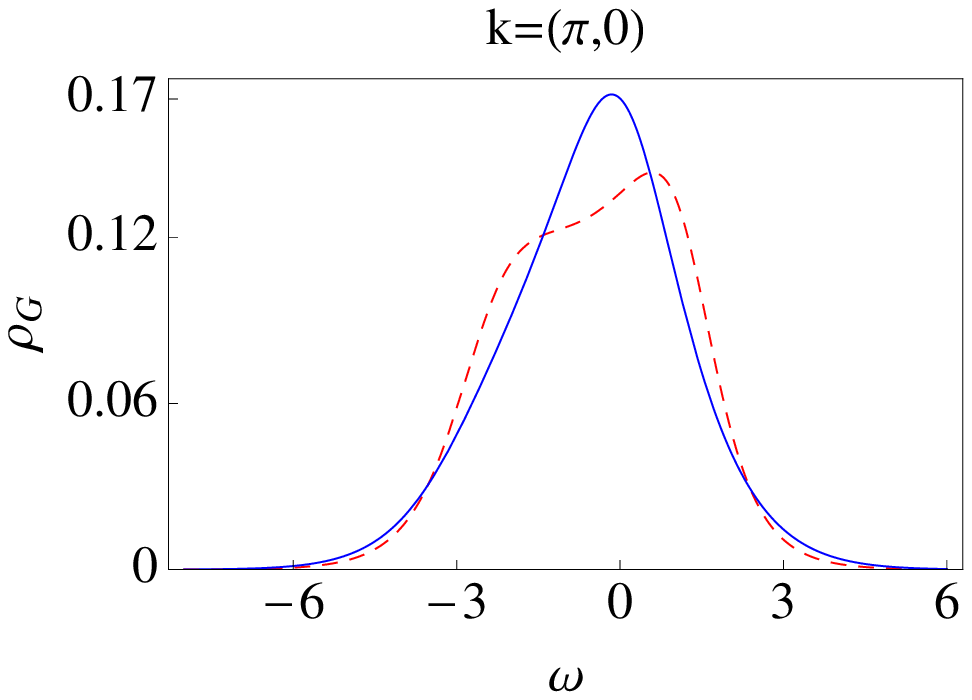}
\includegraphics[width=.68 \columnwidth] {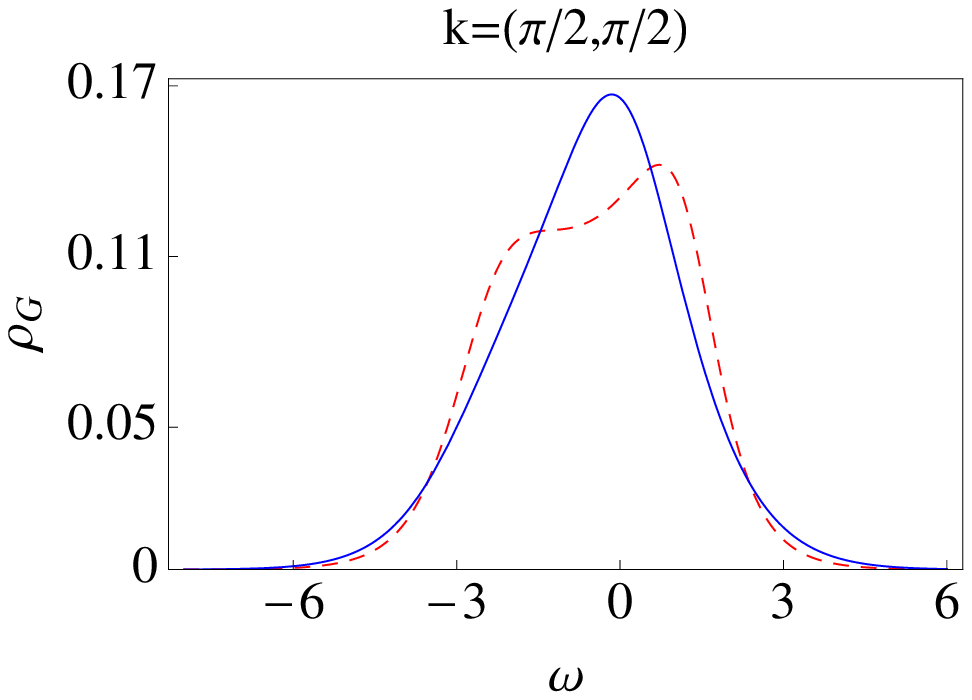}
\includegraphics[width=.68 \columnwidth] {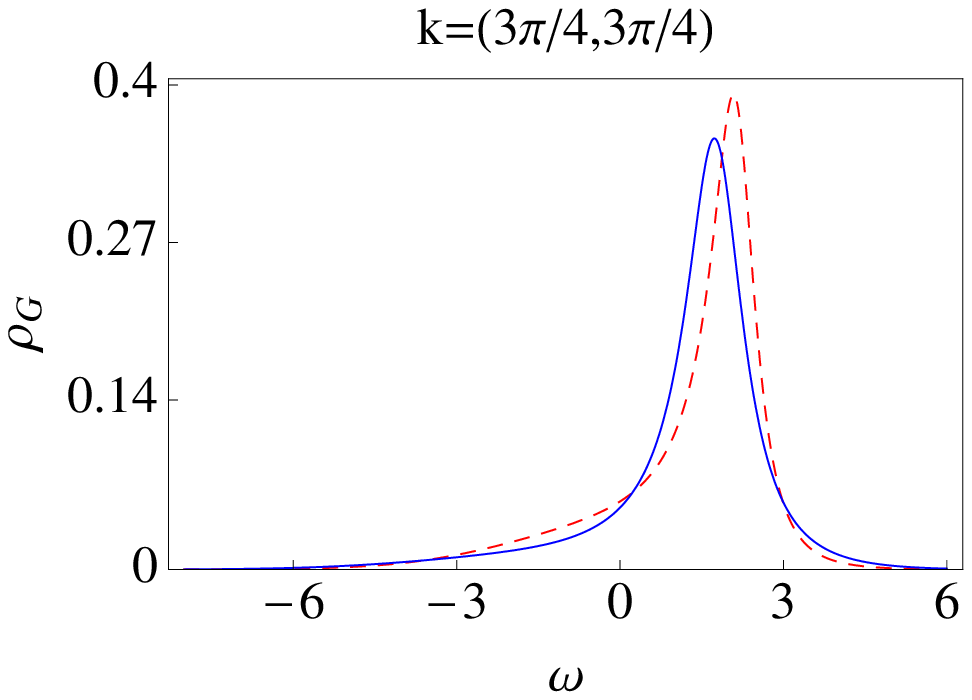}
\includegraphics[width=.68 \columnwidth] {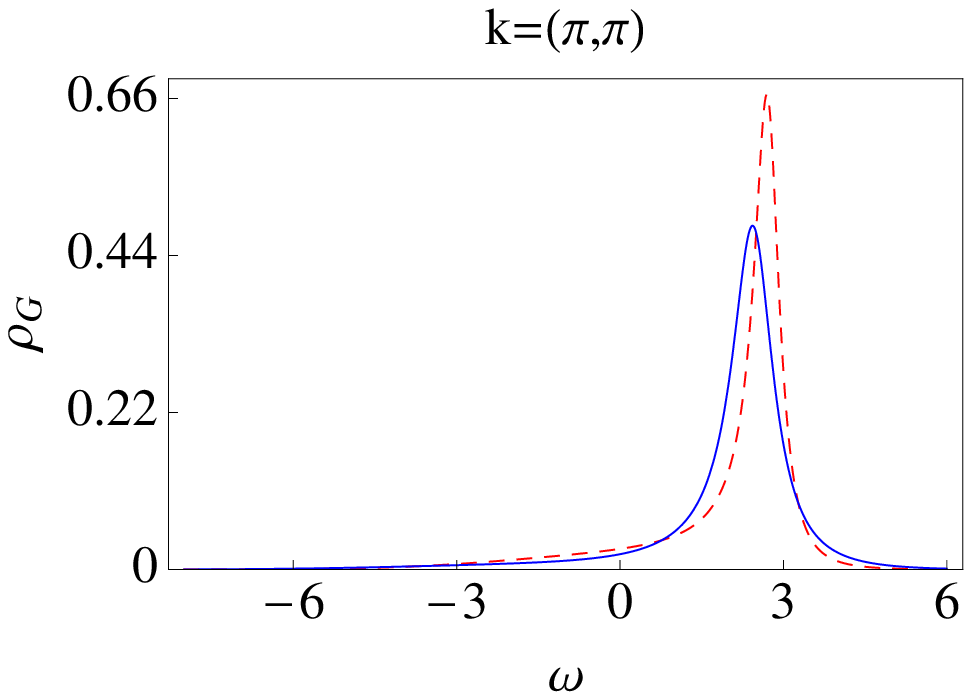}
\end{center}
\caption{The spectral density for the physical Green's function versus
$\omega$ for $T=1.1$ and $n=0.7$. The blue (solid) curve is obtained from 
the Fermi-liquid-type scheme [\disp{rhoSigGauss}] and the red (dashed) curve is obtained from the  
TM scheme [\disp{rhoSigGausssymm}]. The fairly sharp extra peaks 
obtained from the TM scheme,   as compared to the Fermi-liquid scheme, seem to 
be physically unreasonable.  We also note that the spectral functions from ECFL 
found numerically using  the $O(\lambda^2)$ scheme [see Fig. 3(f) of \refdisp{e_khatami_13}] 
find rather broad peaks at high temperatures.}
\label{n7cf}
\end{figure*}

\begin{figure*}[h]
\begin{center}
\includegraphics[width=.68\columnwidth] {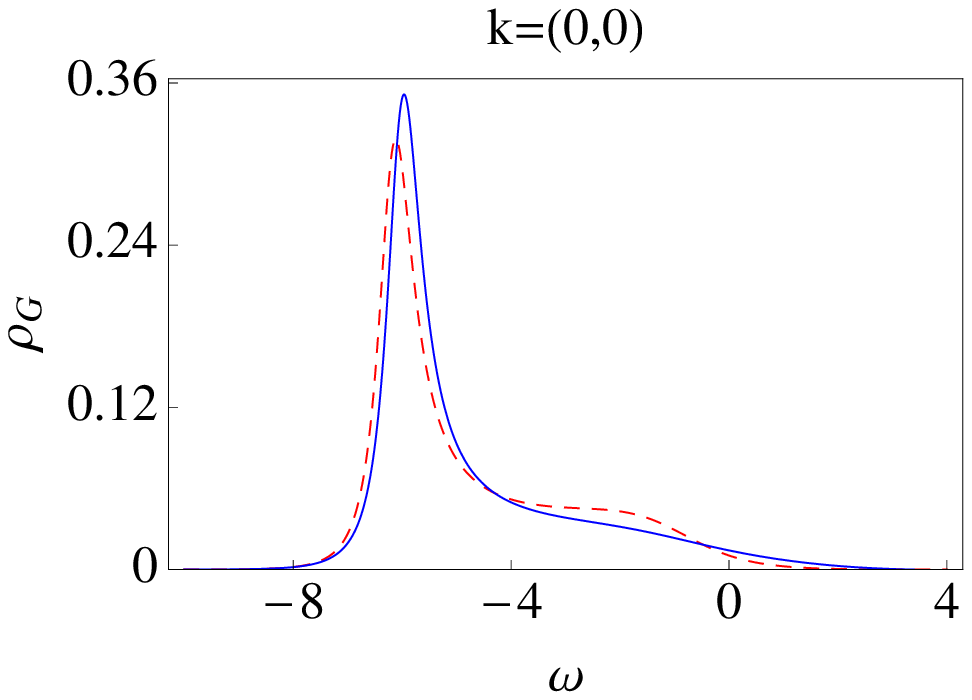}
\includegraphics[width=.68\columnwidth] {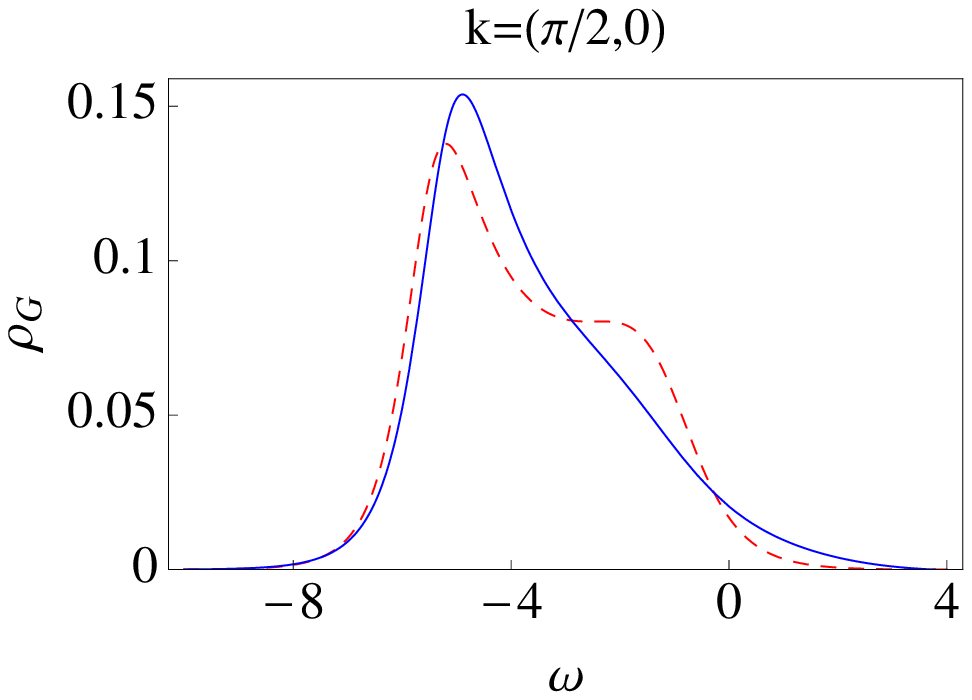}
\includegraphics[width=.68\columnwidth] {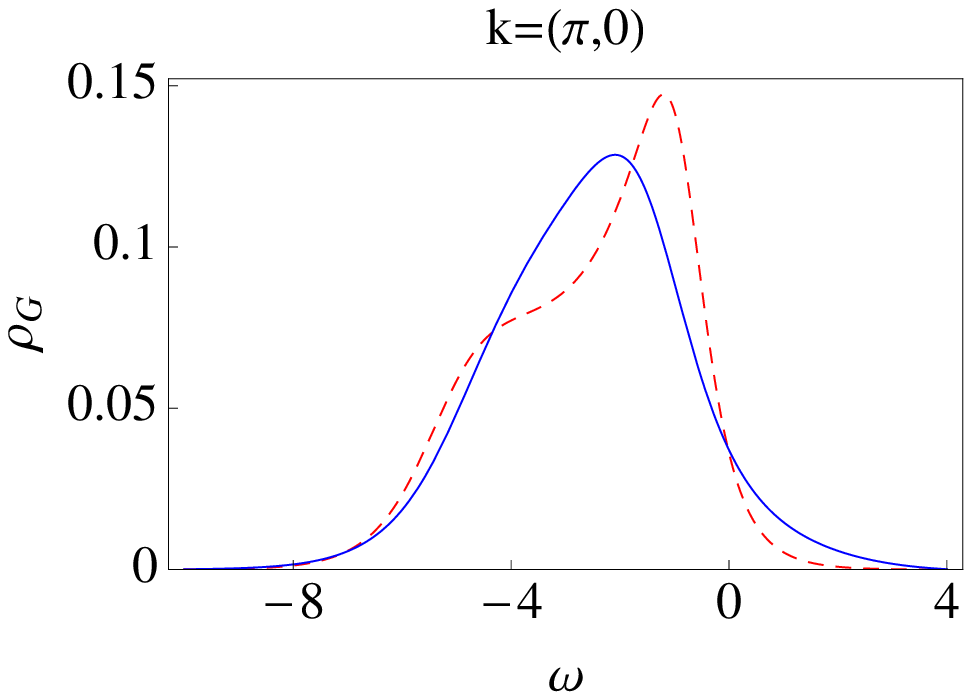}
\includegraphics[width=.68\columnwidth] {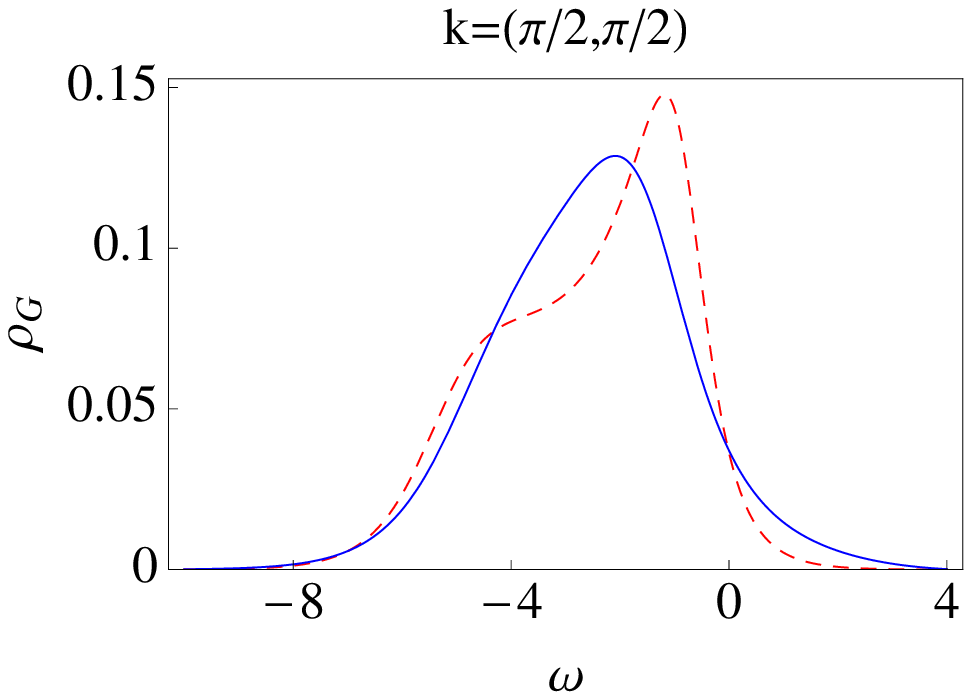}
\includegraphics[width=.68\columnwidth] {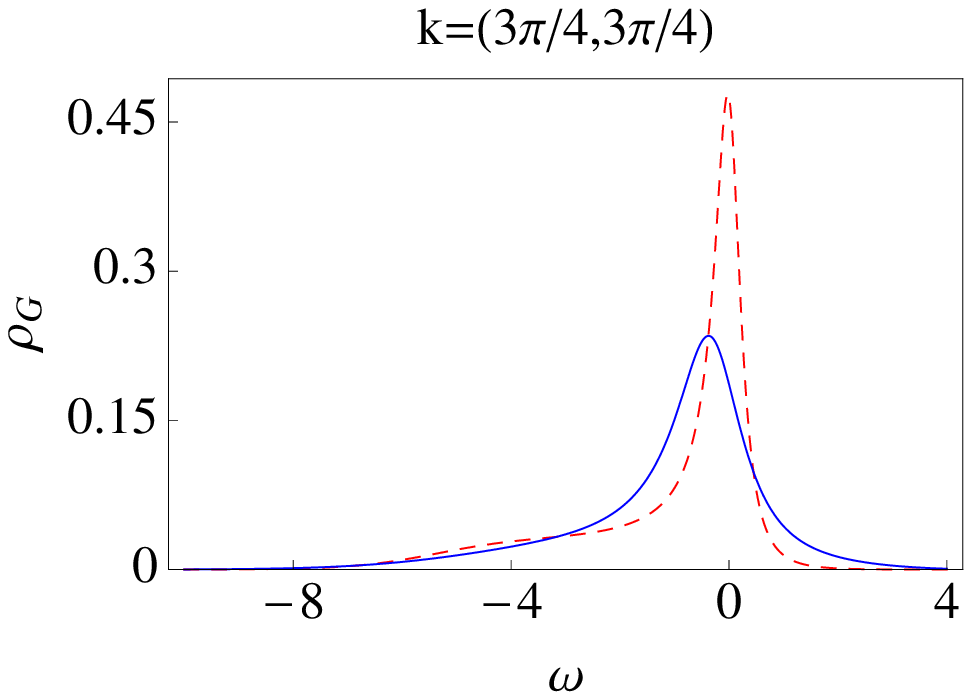}
\includegraphics[width=.68\columnwidth] {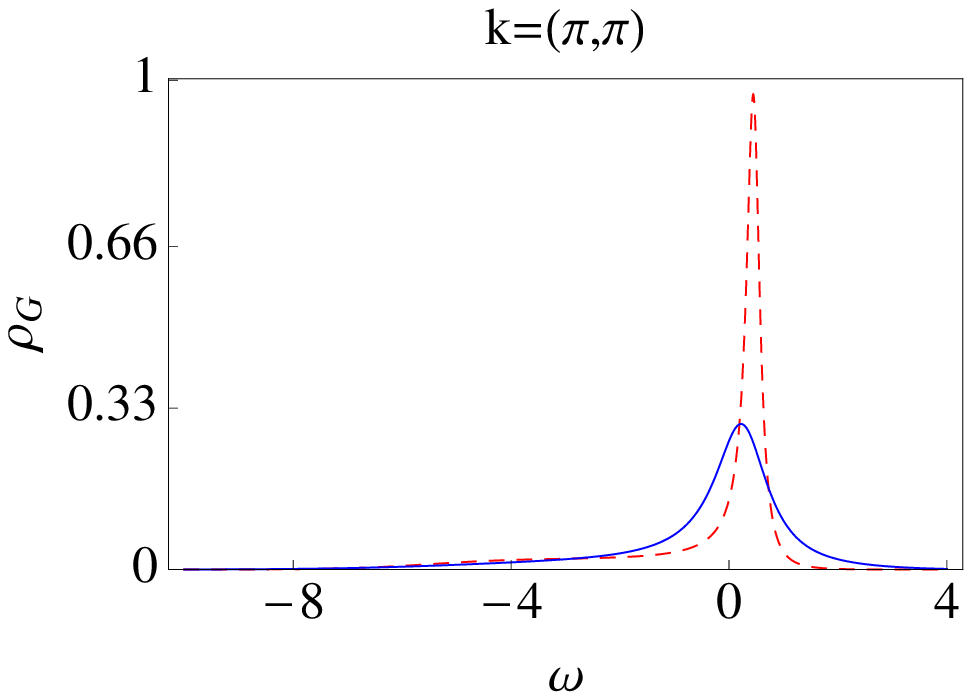}
\end{center}
\caption{The spectral density for the physical Green's function vs
$\omega$ for $T=1.1$ and $n=0.9$. Lines are the same as in Fig.~\ref{n7cf}.}
\label{n9cf}
\end{figure*}

In \refdisp{s_shastry_11}, Shastry establishes the relationship between the continued 
fraction representation of the Green's function [\disp{cfraction}], and a representation 
in terms of an infinite sequence of self-energies with spectral densities 
$\rho_\Sigma^{(n)}(\omega)$, with $n=0,1,\ldots$. For the standard self-energy we omit the superscript
 so that  $\rho_\Sigma^{(0)}(\omega)\equiv \rho_\Sigma(\omega)$. 
This is a particularly convenient 
reformulation of the well-known Mori scheme \cite{Mori} for relaxation processes,
where Laplace transforms over time-dependent correlations are used. In particular, 
denoting $\Sigma_{\infty} \equiv \lim_{z\to\infty}\Sigma(z)$, and recalling that 
$G(z,k) = \frac{a_{\scriptscriptstyle G}}{i\omega + \mu - a_{\scriptscriptstyle G}\, \epsilon_k - \Sigma_\infty -
 \int d\nu \ \frac{\rho_\Sigma(\nu)}{i\omega-\nu}},$
 $b_1 = -a_{\scriptscriptstyle G} 
\, \epsilon_k - \Sigma_{\infty}$, and the standard self-energy is expressed as
\begin{eqnarray}
\hspace{-8mm}\text{$\int  \frac{\rho_\Sigma(\nu-\mu^{(0)})}{z-\nu}
\  d\nu = \frac{a_1}{z+b_2-}\;\; \frac{a_2}{z+b_3-} \;\;\frac{a_3}{z+b_4-} \;\; \frac{a_4}{z},$} \nonumber\\ 
\label{rhoSig}
\end{eqnarray}
where     $ \rho_\Sigma(\omega) \equiv -\frac{1}{\pi} 
{\rm Im}  \  \Sigma(i\omega_n\to\omega+ i\eta)$. Following \cite{s_shastry_11}, we 
identify the constant $a_1\equiv a_\Sigma\equiv\int\rho_{\Sigma}(\nu)d\nu$, $b_2 
\equiv - \Sigma^{(1)}_{\infty}$, and
\begin{eqnarray}
\int  \frac{\rho_\Sigma^{(1)}(\nu-\mu^{(0)})}{z-\nu} \  d\nu =  \frac{a_2}{z+b_3-} 
\;\;\frac{a_3}{z+b_4-} \;\; \frac{a_4}{z}. \nonumber\\ 
\end{eqnarray}
For $l>1$, one has the general formula 
\begin{eqnarray}
 a_l= a_{\Sigma^{(l-1)}}; \;\;\;\;   b_l= -\Sigma_{\infty}^{(l-1)}.
\end{eqnarray}

The Green's function of \disp{cfraction} will lead to a spectral function with a 
small number of well-separated poles and residues. To obtain a continuous shape 
for the spectral function, there are several alternatives. We initially
follow the procedure of Tomita and Mashiyama (TM) \cite{TomitaMashiyama}, which 
is useful in the spin relaxation problems, but does not seem to have features of 
a fermionic self-energy function built into it.  Nevertheless, we try it out in 
view of its simplicity, and  as it provides a counterpoint to our preferred method 
presented next. In the spirit of \refdisp{TomitaMashiyama} , we assume that
\begin{eqnarray}
\rho_\Sigma(\omega-\mu^{(0)}) = A \exp[-\alpha^2(\omega-\omega_0)^2],
\label{rhoSigGausssymm}
\end{eqnarray}
so that the coefficients $A, \alpha, \omega_0$ are fixed using the moments, and 
higher moments are forced to be those of the Gaussian.
Using \disp{rhoSig},
we can solve for $A$, $\alpha$, and $\omega_0$ in terms of $a_1$, $a_2$, and $b_2$.
It is also possible to obtain a continuous spectral function whose moments correctly 
reproduce all of the coefficients in \disp{cfraction} by making the Gaussian 
approximation for the second-level self-energy:
\begin{eqnarray}
\rho_\Sigma^{(2)}(\omega-\mu^{(0)}) = A \exp[-\alpha^2(\omega-\omega_0)^2].
\label{rhoSig2Gauss}
\end{eqnarray}
Then, using the relation,
\begin{equation}
\int  \frac{\rho_\Sigma^{(2)}(\nu-\mu^{(0)})}{z-\nu} \  d\nu =  \frac{a_3}{z+b_4-} \;\; \frac{a_4}{z+\ldots},
\end{equation}
we can solve for $A$, $\alpha$, and $\omega_0$ in terms of $a_3$, $a_4$, and $b_4$. 
However, as shown in \figdisp{compare02} below, this is actually a worse approximation 
as it accentuates an unphysical sharp peak in the TM scheme spectral function.

An alternative scheme for obtaining continuous spectral functions makes use of our 
knowledge of the approximate form of the self-energy as $(T,\omega)\to0$ \cite{ECFLDMFT}:
\begin{eqnarray}
\rho_{\Sigma}(\omega) &=& A (\omega^2 +\pi^2 T^2)\left(1-\frac{\omega}{\Delta}\right)\nonumber\\
&\times&\exp\left[-\frac{\omega^2 +\pi^2 T^2}{\omega_c^2}\right]. 
\label{rhoSigGauss}
\end{eqnarray}
Here, $(\omega^2 +\pi^2 T^2)$ is the standard Fermi-liquid form, $\frac{1}{\Delta}$ 
provides the aforementioned particle-hole asymmetry, and the exponential extrapolates 
the low energy answer to higher energies in a natural way \cite{ecfl}.  Once again, 
we can solve for $A$, $\Delta$, and $\omega_c$ in terms of $a_1$, $a_2$, and $b_2$ 
by using \disp{rhoSig}.

We obtain the spectral function $\rho_G(\omega,k)$
using both \disp{rhoSigGausssymm} and \disp{rhoSigGauss} 
at $T=1.1$ for $n=0.7$ and $n=0.9$ and at various points along the irreducible wedge 
of the Brillouin zone. The spectral functions $\rho_G(\omega,k)$ are plotted in \figdisp{n7cf} 
for $n=0.7$ and in  \figdisp{n9cf} for $n=0.9$.

\section{Summary}
\label{sec:summ}

We present an implementation of the linked-cluster expansion for the Green's function 
of the infinite-$U$ Hubbard model on a computer, which is based on a formalism proposed 
by Metzner~\cite{w_metzner_91}. Using efficient algorithms on parallel computers,
we have carried out the expansion up to the eighth order in terms of the hopping amplitude,
and obtained analytic results for the Green's function and the Dyson-Mori self-energy 
on the square lattice as a function of  momentum and Matsubara frequency at a given 
fixed density. Since the lattice sums for graphs in this approach are evaluated independently 
of their time integrals and spin sums, our implementation paves the way for 
obtaining similar results for other geometries and spatial dimensions. 

To extend the region of
convergence in temperature, we employ \Pd approximations and study several static and 
dynamic quantities. The equation of state exhibits significant deviations from the atomic 
limit starting at relatively high temperatures and reveals interesting trends near $n=0.5$, 
where we find that the chemical potential changes linearly with temperature and remains very close 
to the one in the atomic limit down to the lowest temperatures accessible to us. We also 
find that the change in sign of the derivative of $\mu$ with respect to $T$ at constant density, 
which is proportional to the thermopower in Kelvin's formula, takes place at increasingly 
higher densities due to correlations as the temperature is lowered. The momentum distribution
function also shows significant deviations from free fermions, and becomes more uniform 
across the Brillouin zone as the correlations build up at higher densities. We further study 
dynamic quantities, such as the analog of the quasiparticle fraction in the Matsubara 
frequency space vs temperature, which shows a nonmonotonic dependence on density
at low temperatures, and the lifetime of the quasiparticles at various densities, 
obtained in the series through the first two moments of the electronic spectral functions. 
To make contact with experiments and extend previous results for the spectral functions 
obtained within the ECFL or the dynamical mean-field theory, we calculate them here after 
transforming the Green's function series to continued fractions, or by employing certain forms
for the spectral functions suggested by the ECFL theory. We present our results for densities 
close to half filling at several points in the momentum space.

To benchmark our results from the \Pd approximations for the equation of state at temperatures 
lower than the hopping amplitude, where the direct sums in the series do not converge, and 
to shed more light on the state of the system at those temperatures, we also present results 
from the NLCE up to eleventh order for an equivalent model, i.e., the \tj model with $J=0$. 
We find perfect agreement between the direct sums from the two methods when they converge, and 
that at lower temperatures, the \Pd approximants generally overestimate the density for a 
given chemical potential.
The NLCE results after numerical resummations also help obtain the thermopower vs density at 
a temperature that is not otherwise accessible to the series even after the \Pd approximations.

\section*{Acknowledgments}
This work was supported by DOE under Grant No.~FG02-06ER46319 (B.S.S. and E.P.), 
and by NSF under Grant No.~OCI-0904597 (E.K. and M.R.).

\appendix
\section{RECURSIVE EXPANSION OF CUMULANTS}
\label{app:cmlnt}

In the following, we combine the time and spin variables and denote them by their index only, i.e., 
$C^0_{m}(\tau_1\sigma_1, \dots \tau_m\sigma_m|\tau'_1\sigma'_1, \dots \tau'_m\sigma'_m) 
\to C^0_{m}(1, \dots m|1', \dots m')$. Cumulants are calculated by taking functional derivatives 
of a generating functional with respect to Grassmann variables \cite{w_metzner_91,OrlandNegele}, 
and can be expressed in terms of UGFs. We give explicit expressions for $C^0_{m}$ through $m=3$.
\
\begin{eqnarray}
C^0_1(1|1')&=&G^0_1(1|1'), \nonumber \\
C^0_2(1,2|1',2')&=&G^0_2(1,2|1',2')\nonumber\\
&-&G^0_1(1|1')G^0_1(2|2')\nonumber\\
&+&G^0_1(1|2')G^0_1(2|1'),
\label{Clow}
\end{eqnarray}
\begin{widetext}
\begin{eqnarray}
C^0_3(1,2,3|1',2',3')&=&G^0_3(1,2,3|1',2',3') \nonumber \\
&-&  C^0_2(1,2|1',2')G^0_1(3|3')+C^0_2(1,2|1',3')G^0_1(3|2')-C^0_2(1,2|2',3')G^0_1(3|1')\nonumber\\
&+& C^0_2(1,3|1',2')G^0_1(2|3')+C^0_2(1,3|2',3')G^0_1(2|1')-C^0_2(1,3|1',3')G^0_1(2|2')\nonumber\\
&-& C^0_2(2,3|1',2')G^0_1(1|3')-C^0_2(2,3|2',3')G^0_1(1|1')+C^0_2(2,3|1',3')G^0_1(1|2')\nonumber\\
&-&G^0_1(1|1')G^0_1(2|2')G^0_1(3|3')+G^0_1(1|1')G^0_1(2|3')G^0_1(3|2')+G^0_1(1|2')G^0_1(2|1')G^0_1(3|3')\nonumber\\
&-&G^0_1(1|2')G^0_1(2|3')G^0_1(3|1')+G^0_1(1|3')G^0_1(2|2')G^0_1(3|1')-G^0_1(1|3')G^0_1(2|1')G^0_1(3|2').
\label{C3}
\end{eqnarray}
\end{widetext}
The rule for obtaining the expansion for $C^0_{m}(1, \dots m|1', \dots m') - G^0_{m}(1, \dots m|1', \dots m') $
 is as follows. 
Partition the unprimed integers $1 \dots m$ into at least two sets. Each set in the partition corresponds 
to a cumulant, in which the unprimed numbers in the set are written in ascending order. The primed 
numbers $1' \dots m'$ are then partitioned amongst the cumulants created by the unprimed number 
partitions, and are also written in ascending order. The sign of the term is $(+)$ if the permutation to 
get from primed to unprimed numbers is odd, and $(-)$ if it is even. The sign is due to the Grassmann 
variables in the generating functional, and is ultimately a consequence of the fermionic nature of the 
operators. $C^0_3(1,2,3|1',2',3')$ can be expressed in terms of the UGFs by plugging \disp{Clow} into 
\disp{C3}. In general, $C^0_{m}(1, \dots m|1', \dots m')$ can be obtained in terms of UGFs of equal 
or lower orders by this recursive procedure.

\section{TIME INTEGRALS}
\label{app:TI}

In evaluating the time integrals, we use the following general 
result for the time integral of a product of 
step functions in terms of a series of ordered internal times, $\tau_i$,
over which the integrals are taken, and a fixed external time, $\tau$:

\begin{widetext}
\begin{eqnarray}
\int_0^{\beta}&d\tau_n& \int_0^{\beta}d \tau_{n-1} \dots \int_0^{\beta}d \tau_2 \int_0^{\beta}d \tau_1 
\Theta(\tau_n-\tau_{n-1})\Theta(\tau_{n-1}-\tau_{n-2})\dots
\Theta(\tau_{m+1}-\tau) \Theta(\tau-\tau_m) \dots
\Theta(\tau_3-\tau_2) \Theta(\tau_2-\tau_1)\nonumber \\
&=&\frac{\tau^m(\beta-\tau)^{n-m}}{m!(n-m)!}.
\end{eqnarray}
\end{widetext}


\begin{thebibliography}{99}
\bibitem{self-consistent} There have been a number of efforts to perform a 
strong-coupling expansion for the Green's function in a self-consistent manner, i.e., by 
resuming an infinite subclass of diagrams \cite{l_craco, Pairault}.
\bibitem{compare} This series has also been calculated by hand through fourth order  
in \refdisp{Edward} using a completely independent method. The results of the two series match exactly.
\bibitem{note2}
\bibinfo{note}{Note that when $U=\infty$, the creation and annihilation operators in Eq.~\ref{eq:g}
become Gutzwiller projected, and hence, follow different anticommutation relations than those 
valid for canonical fermion operators.}
\bibitem{note1}
\bibinfo{note}{Higher orders for the square lattice or other geometries are available upon request.}
\end{thebibliography}
\end{document}